\documentclass[letterpaper,twocolumn,10pt]{article}
\usepackage{usenix2019_v3}

\usepackage{tikz}
\usepackage{amsmath}
\usepackage{subfig}
\usepackage{multirow}
\usepackage{filecontents}
\usepackage{graphicx}
\usepackage{epstopdf}
\usepackage{booktabs}
\usepackage{algorithmicx}
\usepackage[Algorithm,ruled]{algorithm}

\usepackage{authblk}
\makeatletter
\renewcommand\AB@affilsepx{, \protect\Affilfont}
\makeatother

\begin{document}

\date{}

\title{\Large \bf EaTVul: ChatGPT-based Evasion Attack Against Software Vulnerability Detection}


\author[1, 2]{\rm Shigang Liu}
\author[2]{\rm Di Cao}
\author[3]{\rm Junae Kim}
\author[3]{\rm Tamas Abraham}
\author[3]{\rm Paul Montague}
\author[1]{\rm Seyit Camtepe}
\author[2]{\rm Jun Zhang}
\author[2]{\rm Yang Xiang} 

\affil[1]{\rm CSIRO's Data61}

\affil[2]{\rm Swinburne University of Technology}

\affil[3]{\rm DST Group, Australia}

%

\maketitle
\pagestyle{empty}

\begin{abstract}
Recently, deep learning has demonstrated promising results in enhancing the accuracy of vulnerability detection and identifying vulnerabilities in software. However, these techniques are still vulnerable to attacks. Adversarial examples can exploit vulnerabilities within deep neural networks, posing a significant threat to system security.
This study showcases the susceptibility of deep learning models to adversarial attacks, which can achieve  100\% attack success rate (refer to Table \ref{table:three_target_model}). The proposed method, EaTVul, encompasses six stages: identification of important samples using support vector machines, identification of important features using the attention mechanism, generation of adversarial data based on these features using ChatGPT, preparation of an adversarial attack pool, selection of seed data using a fuzzy genetic algorithm, and the execution of an evasion attack. Extensive experiments demonstrate the effectiveness of EaTVul, achieving an attack success rate of more than 83\% when the snippet size is greater than 2. Furthermore, in most cases with a snippet size of 4, EaTVul achieves a 100\% attack success rate.
The findings of this research emphasize the necessity of robust defenses against adversarial attacks in software vulnerability detection.
\end{abstract}

\section{Introduction}
Software vulnerability detection systems play a crucial role in safeguarding computer systems and networks. Deep neural networks have made significant advancements in this field, as demonstrated by recent algorithms ~\cite{ghaffarian2017software}, ~\cite{le2022survey}. 
For instance, Lin et al. \cite{lin2017poster} extract high-level function representations from the abstract syntax tree (AST) to detect function-level vulnerabilities across projects.
Feng et al. \cite{feng2020efficient} propose a method utilizing the AST to extract syntax features and minimize data redundancy. Yang et al. \cite{yang2021asteria} leverage a deep learning-based  method using the Tree-LSTM network to assess the semantic equivalence of functions across platforms. Fu and Tantithamthavorn \cite{fu2022linevul} present LineVul, a Transformer-based approach for line-level vulnerability prediction. However, it is crucial to acknowledge that these systems can be susceptible to attacks, which can compromise overall security ~\cite{rosenberg2021adversarial}.

Recent studies have revealed that adversarial attacks can exploit vulnerabilities in software vulnerability detection techniques that use machine learning, particularly deep learning \cite{schuster2021you, yefet2020adversarial}. 
Zhang et al. \cite{zhang2020generating} proposed the Metropolis-Hastings Modifier algorithm to generate adversarial samples for attacking machine learning-based software vulnerability detection systems. Ramakrishnan and Albarghoutthi \cite{ramakrishnan2022backdoors} investigated the feasibility of backdoor attacks on deep learning-based techniques used in software vulnerability detection systems. Henkel et al. \cite{ramakrishnan2022semantic} assessed the current architectures of machine learning-based software vulnerability detection. However, these studies are still in the early stages of exploring adversarial attacks in machine learning-based techniques, 
and the security issues of these techniques have not been thoroughly evaluated. For example, the defense against the scenario of adversarial attacks has not been considered in almost all machine/deep learning-based software vulnerability detection systems \cite{lin2020software, le2022survey, Luo0WTXZLL23, sendner2023smarter, christou2023ivysyn, pearce2023examining, mirsky2023vulchecker}. The adversarial attacks hold significant importance as they allow attackers to modify their samples (e.g., vulnerable samples) to bypass the prediction model. By manipulating the input data, adversaries can deceive the model into making incorrect predictions (e.g., vulnerable samples predicted as non-vulnerable), compromising the overall security of the system. As the number of hackers has grown [3], there is a strong demand to evaluate the security of software analysis techniques. 
%
%
Therefore, this motivates us to conduct fundamental research on evasion attacks to have a thorough understanding of the security issues of machine learning-based software vulnerability detection techniques.

\begin{figure*}[!tb]\centering
	\subfloat[Vulnerable samples predicted as vulnerable \label{Fig.ad1}]
	{\includegraphics[width=0.48\textwidth]{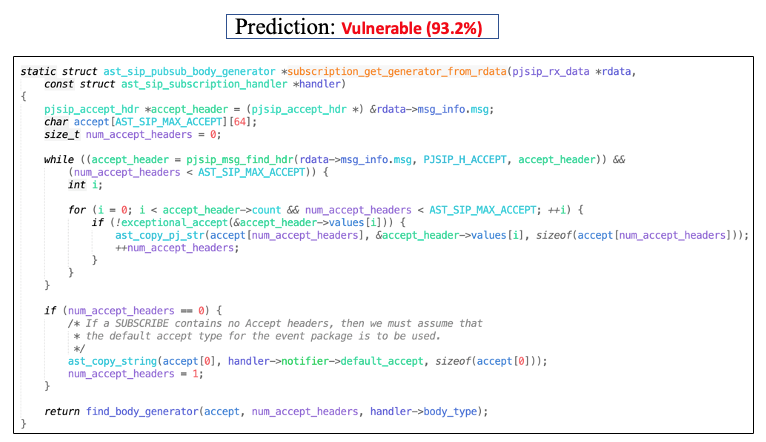}}
	\subfloat[Vulnerable samples predicted as non-vulnerable  \label{Fig.ad2}]
	{\includegraphics[width=0.53\textwidth]{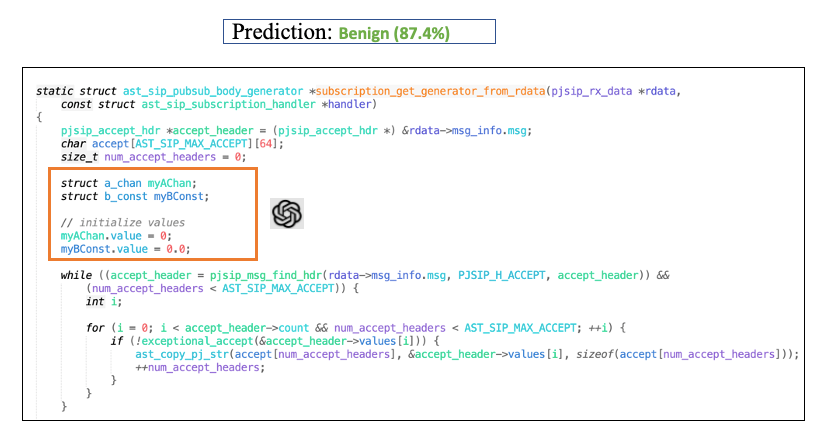}}
	\caption{This figure shows that a vulnerable sample can be easily bypassed by adding a precisely crafted piece of adversarial data. Left: a vulnerable function predicted as vulnerable with a high probability of 93.2\%; Right: the same vulnerable function predicted as non-vulnerable with a high probability of 87.4\% after adding an optimized adversarial data generated by EaTVul. 
	}
	\label{Fig.ad}
\end{figure*}

In this paper, we propose EaTVul (\underline{E}vasion \underline{At}tack Against Software \underline{Vul}nerability Detection), an automatic attack strategy from the perspective of an attacker. We assume no knowledge of the target model and cannot manipulate the training data. Our experiments demonstrate that EaTVul achieves an attack success rate of more than 83\% when the snippet size is greater than 2 and 100\% for most cases with a snippet size of 4.
At a high level, our approach involves using SVMs (support vector machines) to identify important vectors in non-vulnerable samples. We then employ an attention mechanism to highlight important features contributing to predictions in non-vulnerable samples. Using chatGPT, we generate adversarial data based on these important features and prepare a preserved attack pool. We use a fuzzy genetic algorithm to automatically optimize the selection of seed adversarial data and insert it into vulnerable samples, aiming to make them predicted as non-vulnerable.
Figure 1 illustrates how a vulnerable function can be easily bypassed by adding precisely crafted adversarial data generated by EaTVul. The prediction changes from 93.2\% vulnerable to 87.4\% non-vulnerable.
In summary, our main contributions are as follows:

\begin{itemize}   
   \item We propose a novel evasion attack approach, named EatVul, which produces code to evade ML-based vulnerability detectors. To achieve this, EatVul first employs SVM to identify the most important non-vulnerable samples, which will be used as important samples to identify the important features. Then, it utilizes the attention mechanism to identify the important features that have the most significant contribution to the prediction, based on the important samples identified by SVM. Afterwards, it uses ChatGPT to generate adversarial data based on the important features identified by the attention model. Finally, EatVul uses the improved fuzzy genetic algorithm (FGA) to select the optimal seed adversarial data for launching an evasion attack against machine learning-based software vulnerability detection systems.
   \item We have conducted an evaluation of EatVul against state-of-the-art baselines, and the experimental results demonstrated that our scheme achieved a 100\% success rate for most cases (refer to Table.\ref{table:three_target_model} with Sni.-4). Our study presents significant findings to the software security community. 
   \item We have made our proposed system, EatVul, available to the research community. Furthermore, we have published the datasets and code to encourage others to contribute to adversarial learning. We hope that this system will raise the attention for the security community to have good understanding the security issues of machine learning/deep learning-based systems for software security and to develop further defense strategies. The datasets and code are available at  \url{https://github.com/wolong3385/EatVul-Resources}.
\end{itemize}

\section{Related Work}
\label{section: Related Work}
In this section, we will only review works closely related to this study. For more comprehensive information about adversarial machine learning in other research areas such as computer vision, natural language processing, and cybersecurity, please refer to \cite{maiorca2019towards, rosenberg2021adversarial, machado2021adversarial, li2021conaml}.

Recent research has highlighted the vulnerability of machine learning, particularly deep learning, techniques to adversarial attacks in the field of software vulnerability detection \cite{schuster2021you, yefet2020adversarial}. Zhang et al. \cite{zhang2020generating} 
introduced the Metropolis-Hastings Modifier algorithm to generate adversarial samples specifically for attacking machine learning-based software vulnerability detection systems. 
Zeng et al. \cite{zeng2021} developed OpenAttack, an open-source toolkit for textual adversarial attacking with unique strengths in supporting all attack types, multilinguality, and parallel processing.
Yang et al. \cite{yang2022natural} further improved the strategy using a greedy and genetic algorithm with a focus on semantic preservation. 
%
Srikant et al. \cite{Srikant2021GeneratingAC} further combined site-selection and perturbation-choice into a joint mathematical problem, proposing a set of first-order optimization algorithms to solve the formulation. Their approach has demonstrated a 1.5x increase in attacking performance over previous adversarial generation methods \cite{yang2022natural}, which we use as one of our baseline models. 
Yu et al. \cite{yu2023advulcode} presented AdVulCode, the first DL-based adversarial example generation method for vulnerability detection models, demonstrating its effectiveness through controlled perturbation and an improved MCTS search algorithm. The recently published code obfuscation tool, Milo, proposed by Song et al. \cite{song2023milo}, focuses on an elevated abstraction tier, applying the transformation to the parsed abstract syntax trees (AST). 

Ramakrishnan and Albarghoutthi conducted a study on the potential of backdoor attacks targeting deep learning-based techniques used in software vulnerability detection based on source code \cite{ramakrishnan2022backdoors}. The authors introduced additional data points containing triggers into the original training dataset, and experimental results revealed that code2seq and seq2seq-based techniques are susceptible to backdoor attacks. 
Henkel et al. \cite{ramakrishnan2022semantic} evaluated state-of-the-art machine learning-based architectures for software vulnerability detection and observed that code2seq surprisingly exhibits vulnerability to adversarial attacks. 
Zhou et al. \cite{zhou2022adversarial} investigated the robustness of deep neural networks (DNNs) in generating code comments and proposed ACCENT, an identifier substitution approach that generates adversarial data. These snippets maintain syntactic correctness and semantic similarity to the original code but can mislead DNNs into producing irrelevant comments. 

\begin{figure*}[!th]
\centering
\includegraphics[scale=0.45]{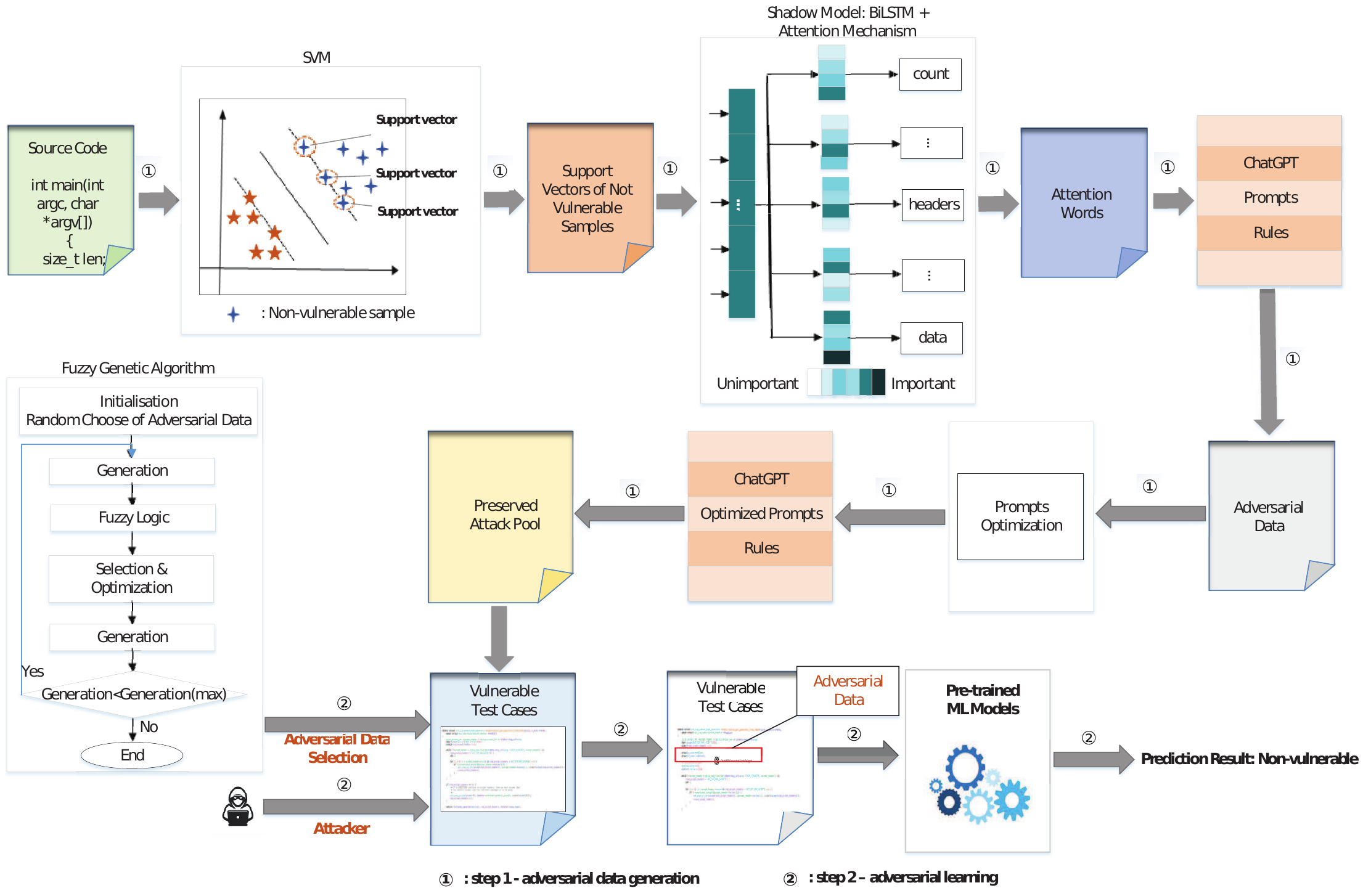}
\caption{The framework of EaTVul.}
\label{Fig:FW}
\end{figure*}

Moreover, 
efforts have been made to assess the robustness of natural language models. For instance, TextFooler \cite{jin2020bert} utilized synonyms of selected keywords combined with part-of-speech (POS) tagging information to replace original words, with the aim of minimizing alterations. BERT-Attack \cite{li-etal-2020-bert-attack} and BAE \cite{garg-ramakrishnan-2020-bae} employed pre-trained masked language models to generate more fluent and natural tokens consistent with contexts. In an effort to enhance the stealthiness of attacks, Yang et al. \cite{yang2020greedy} proposed a greedy search strategy with replacement operations at the character level, applicable to various state-of-the-art text classification models. However, these endeavors were found to be inadequate for programming languages, given their more structural nature and functional semantics compared to natural language.


To the best of our knowledge, few works focus on assessing the robustness of these deep-learning methods. Unlike prior works in text-based adversarial learning \cite{jin2020bert, li-etal-2020-bert-attack, garg-ramakrishnan-2020-bae, yang2020greedy}, this paper focuses on the naturalness of adversarial statements while preserving normal execution by inserting adversarial data. Notably, this work is different from adversarial malicious learning. Generally speaking, malware detection using machine learning usually based on binaries rather than source code \cite{kolosnjaji2018adversarial}.
Therefore, a thorough discussion of adversarial malware is beyond the topic of this paper. For more information, please refer to \cite{maiorca2019towards, li2021arms, yan2022survey}.

\section{Overview of the EaTVul}
This section introduces EaTVul, an automated system designed to attack machine learning-based software vulnerability detection systems. Figure \ref{Fig:FW} provides an overview of the proposed EaTVul, which consists of two main phases: adversarial data generation (\textcircled{1}) and adversarial learning (\textcircled{2}).

In the first place, we train a surrogate model based on BiLSTM with an attention mechanism. To generate the adversarial data, several stages are involved in Phase 1. First, we identify important non-vulnerable samples using SVM. Then, we retrieve the averaged attention scores from the attention layer to identify the key features that contribute significantly to the prediction.
These important features serve as inputs to ChatGPT, which generates adversarial data. The generated adversarial data will then be further reviewed and optimized, and ChatGPT is used again to regenerate the adversarial data. The optimized adversarial data is then added to the preserved attack pool. In Phase 2, our goal is to bypass a machine learning-based software vulnerability detection system using a vulnerable sample. To achieve this, we utilize a fuzzy genetic algorithm to select the best seed data, which is added to the vulnerable test case. The expectation is that the modified vulnerable test case will be predicted as non-vulnerable with a high probability. Details regarding the input/output of each step will be discussed in Section \ref{section:ADG}.


In this study, the attacker's capability, knowledge, and goal are as follows: \textbf{Attacker's capability.}  The attacker is capable of perturbing the test queries given as input to pre-trained vulnerability detectors to generate adversarial samples. We follow the existing paradigm for generating adversarial examples in programming languages \cite{yang2022natural} and allow for two types of perturbations for the input code sequence: (i) token-level perturbations (for instance, variable renaming) and (ii) statement-level perturbations (for example, dead code insertion). To maintain the stealthiness and functionality of the perturbated code samples, we choose statement-level modification in this work. Specifically, the attacker is allowed to insert a certain number of non-functional statements in arbitrary locations.
\textbf{Attacker's Knowledge.} Attacker’s Knowledge. In the context of this study, we employ a conventional black-box framework for deep-learning-based vulnerability detection methods. Here, we presume that attackers do not have access to the architecture and parameters of target models; they are restricted to querying the deployed vulnerability detection model solely with input code sequences, receiving corresponding output probabilities or predictions. However, attackers have the capability to collect all open-source resources, including vulnerability information from NVD, to train their surrogate prediction model. Since practitioners typically utilize public vulnerability repositories to construct their training datasets, real-world vulnerable samples are limited with respect to vulnerability types. Therefore, we assume there will be overlap between the training data collected by attackers and practitioners. We maintain that this black-box setting is practical, unlike white-box or grey-box scenarios, where attackers are assumed to have access to all or part of the aforementioned facets.
\textbf{Attacker's Goal.} When presented with an input code sequence or program representations from a vulnerable program, the objective of the attacker is to deceive the targeted vulnerability detection tools through imperceptible modifications to the inputs. Importantly, within the framework delineated in this paper, we formulate an additional requirement: the inserted adversarial code snippets must refrain from adversely affecting the regular execution of the code samples under examination.


\begin{figure}[!t]
\centering
\includegraphics[scale=0.73]{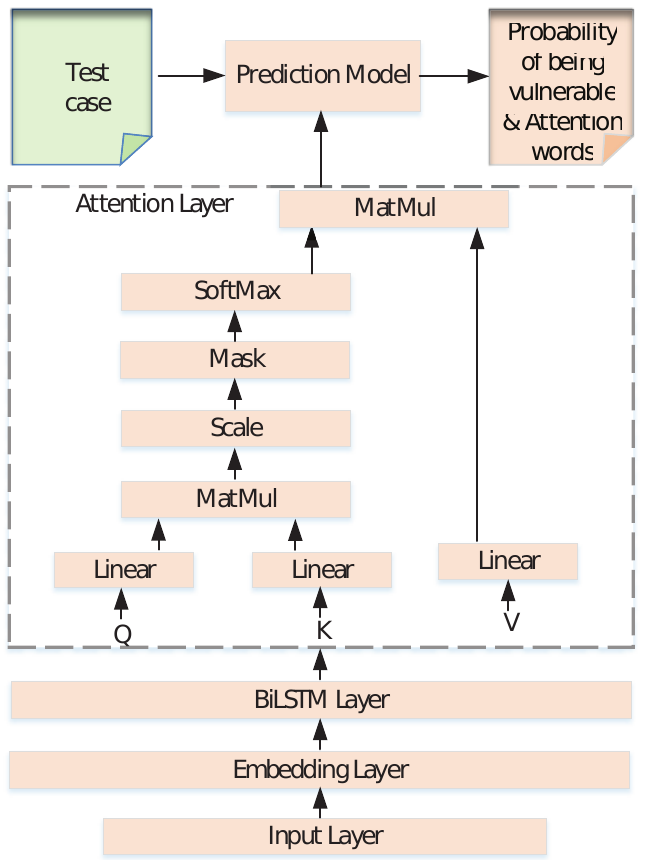}
\caption{Framework of feature learning and attention mechanism.}
\label{fig:attention}
\end{figure}

\subsection{Adversarial Data Generation}
\label{section:ADG}
In this study of software vulnerability detection, adversarial data generation involves adding adversarial data in the vulnerable samples that are intentionally designed to deceive machine learning algorithms into making incorrect predictions or decisions. The goal of generating adversarial data is to identify weaknesses in the machine learning models used to detect vulnerabilities in software. Adversarial data should meet the following requirements:
1) It should include all the important features identified by the attention mechanism;
2) The adversarial data must maintain the code's functionality and should not introduce any syntactic errors or alter its operation;
3) The size of the adversarial data should be limited to less than 8 lines in this study to enhance its concealment and make it challenging to detect.
To meet these requirements, we employ ChatGPT to generate adversarial data while considering all the important features (requirement 1). Subsequently, the raw adversarial data generated will undergo further optimization through prompts optimization (requirement 2) and re-generation using ChatGPT (requirement 3). The following context will provide a detailed explanation of these steps.

\subsubsection{Important Samples Identification using SVM}
A well-known fact is that not every sample contributes equally to the prediction. In this paper, our objective is to select the most important non-vulnerable samples and then identify the features that contribute the most to the prediction. Hence, a layered approach is essential to pick the most critical code samples and identify the most important features within those code samples. Therefore, we employ SVM, which is a low cost way to identify the important code segments (i.e., important non-vulnerable samples, which are the data points that lie closest to the decision boundary of the machine learning model) \cite{liu2018data}. 
By analyzing the properties of these support vectors, it is possible to gain insights into the most important features or characteristics of the data that the model is using to make its predictions. Once the most important samples have been identified, an attacker can then use this information to generate new data points that mimic these samples in order to manipulate the system's behavior. This could involve adding or modifying features in a way that is designed to fool the machine learning model into making incorrect predictions or decisions.

Assume a training software dataset of   of $ n $ samples, $D= \{(x_{i}, y_{i})\}_{i}^{n} $, a soft margin SVM learns the weights $w$ and bias $b$ by solving the following convex QP optimization problem \cite{pisner2020support}:
\begin{equation}\label{E1}  
minL(w,b,\xi) = min\frac{1}{2}\left \| w \right \|^{2}+C\sum_{i}^{n}{\xi_{i}}      
\end{equation}
such that
\begin{equation*}\label{E2}  
y_{i} (w^{T} { x}_{i}+b) \geq 1-\xi_{i},  
\end{equation*}
and
\begin{equation*}\label{E3}  
\xi_{i}\geq 0, \ \  i=1,2,\cdots, n.
\end{equation*}
In the context of machine learning, let $D$ represent the training dataset. The optimization process aims to maximize the margin by minimizing the term $\frac{1}{2}\left|w\right|^2$, where $w$ denotes the weight vector. The hinge loss, represented by the variable $\xi_i$, quantifies the classification error. The regularization parameter $C$ controls the balance between minimizing the classification error on the training data and maximizing the margin.

In this stage, the input consists of high-level program representations from the last layer of the surrogate model, denoted as $D = {(x_i, y_i)}_i^n$, where $y_i$ is the label of input samples. The output of interest in this stage is the set of support vectors. Specifically, we focus on evasion attacks given a vulnerable sample, so we only select the support vectors from the non-vulnerable class as the important samples. These support vectors are represented as $sv = {(sv_1, sv_2, ..., sv_l)}$, and they will be used in the subsequent subsection discussed in Section \ref{feature_identification}.

\begin{figure}[!t]
\centering
\includegraphics[scale=0.62]{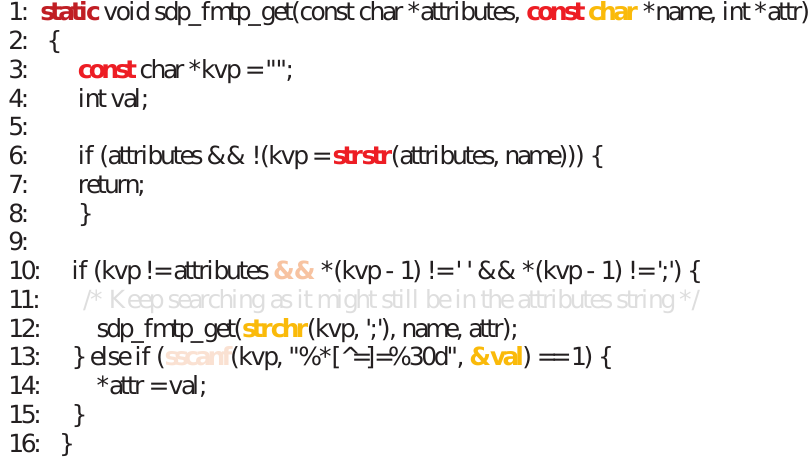}
\caption{Important features identified by attention mechanism display. The importance decreases from red to yellow.}
\label{fig:atw}
\end{figure}

\subsubsection{Important Feature Identification}
\label{feature_identification}
Identifying the most important features is crucial for launching effective adversarial attacks. Adversaries aim to manipulate a system's decision-making process by introducing carefully crafted adversarial examples that resemble normal samples but are misclassified by the system. 
In this paper, we employ BiLSTM (bidirectional long short-term memory) with the attention mechanism \cite{liu2019cyber} as the surrogate model and identify the most important features using the average attention scores. These features will be further utilized to generate adversarial data. In this step, we adopt the indices set from the previous stage and retrieve the weights from the attention layer. Further, the averaged attention score will be projected to the tokens in the statements. To maintain the stealthiness, we exclude low-frequency or unusual user-defined terms unique to a singular project. The outcome of this stage comprises a corpus of candidate features deemed significant.

Figure \ref{fig:attention} illustrates the general overview of the surrogate model. The input data, which is the source program, passes through several layers including the embedding layer, BiLSTM layer, and the self-attention layer. The output of the attention layer is the prediction model, which provides the probability of vulnerability and the attention words for a given test case. The succeeding component of the attention layer is the prediction module, which provides the probability of vulnerability. Figure \ref{fig:atw} showcases an example of the important features identified by the attention mechanism, with the importance depicted through a gradient from red to yellow. In this example, the identified features include \emph{static, const, strstr, strchr, val, sscanf}. These features will be incorporated into the generation of adversarial data.

The purpose of the self-attention mechanism is to capture the dependencies and relationships between different words in a sequence of source programs. It enables the model to assign distinct weights or attention scores to each word based on its relevance to other words in the sequence. By learning these attention weights, the model can effectively focus on important and relevant words, allowing it to capture long-range dependencies. 
Motivated by the previous study \cite{vaswani2017attention}, the softmax attention function in this study is as follows: 
\begin{equation}\label{EA}  
attention (Q,K,V)=softmax(\frac{QK^{T}}{\sqrt{d_{k}}})V  
\end{equation}
where $Q$, $K$, $V$ are matrices of the sets of queries, keys, and values, $d_{k}$ means the dimension. 

%
%
\begin{figure}[!tb]
\centering
\label{Fig.C}
	\subfloat[Raw data generated by ChatGPT \label{Fig.C1}]
	{\includegraphics[width=0.47\textwidth]{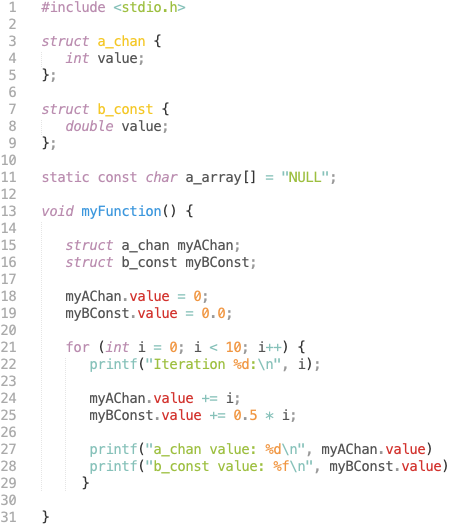}} \\
	\subfloat[Further optimized data by ChatGPT  \label{Fig.C2}]
	{\includegraphics[width=0.44\textwidth]{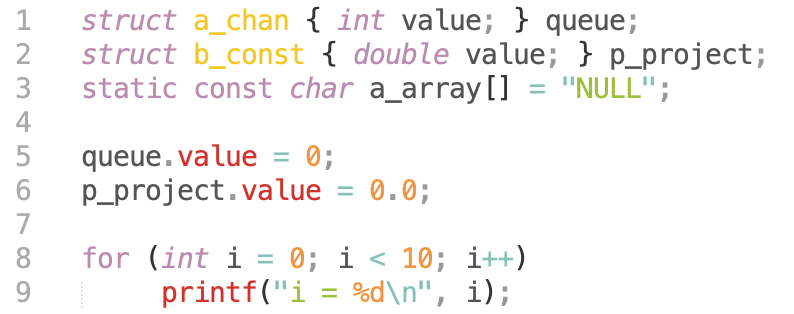}}
	\caption{ {Raw adversarial data and optimized adversarial data generated by ChatGPT.
The top figure contains more than 25 lines of code, while the bottom one displays a more concise version with no more than 8 lines. 
	}}
	\label{fig:raw_op}
\end{figure}

\subsubsection{Adversarial Data Generation using ChatGPT}
\label{chatgpt}
%
Generative AI has the potential to revolutionize various aspects of our lives, with chatbots being one of the most popular implementations. ChatGPT has been extensively used and tested in different domains, showcasing its remarkable capabilities. In our work, we digest the candidate set of significant features and leverage ChatGPT as the code generation tool to generate adversarial data. In comparison to the predefined templates, the code snippets exhibit a higher level of fluency. 

\begin{table}[!tb]
\centering
\small
\caption{Categories of Keywords in C programming Language}
\label{table:category}
\begin{tabular}{ll}
\hline
Category          & Features                                                                                              \\ \hline
Data Type         & `int', `float', `double', `char', etc.                                                                \\ \hline
Control Statement & \begin{tabular}[c]{@{}l@{}}`if-else', `switch-case', `for',  \\ `while', `do-while', etc.\end{tabular} \\ \hline
Storage Classes & \begin{tabular}[c]{@{}l@{}}`auto', `extern', `static', \\ `register', etc.\end{tabular} \\ \hline
Input-Output      & `printf', `scanf', etc.                                                                               \\ \hline
Miscellancous     & \begin{tabular}[c]{@{}l@{}}`sizeof', `return', `break', `typeof', \\ `continue', etc.\end{tabular}    \\ \hline
\end{tabular}
\end{table}

\begin{algorithm*}[!tb]
\centering
\caption{Adversarial Sample Generation using Fuzzy Genetic Algorithm}
\label{table:FAG}
\begin{tabular}{@{}l@{}}
\toprule
\textbf{Input:} \textit{$M$} (DL-based vulnerability detector); \textit{$D_{i}$} (a group of vulnerable programs); \textit{$S_{i}$} (the set of statements for perturbations \\  in $D_{i}$); \textit{$K$} (the number of fuzzy clusters)                                       \\ \hline
\textbf{Output:} A set of optimized adversarial data                                                                           \\ \hline
1: Population \textit{$T$} $\leftarrow$ init($s_{j}$), where $s_{j}$ is the statement in $S_{j}$ ;
                                                        \\ \hline
2: Score set $R(s_{j})$(refer to Equation \ref{ES})$ \leftarrow$ init (score($D_{i} \otimes s_{j}$) ), where $\otimes$ represents insertion, $s_{j}$ is the statement snippet in $S_{j}$.                                                                                 \\ \hline
3: Centroid set $C$ $\leftarrow$ init ($c_{k}$), where $c_{k}$ are randomly generated centroid ranging from $[0, 1)$;                                                                                     \\ \hline
4: Fuzzy cluster label set $L$ $\leftarrow$ init(cluster label $l_{jk}$), where $l_{jk}$ are randomly assigned labels;                                                                                             \\ \hline
5: \textbf{while} there is label change for the element in $L$ or distance perturbations \textbf{do}                                                                                       \\ \hline
6: \qquad \textbf{for} t $\leftarrow$ 1$.... j$ \textbf{do}                                                                                             \\ \hline
7: \qquad \quad calculate distance $d_{k}$ to $k$ clusters using  \textit{Equation} \ref{EL1} \& \textit{Equation} \ref{EL2};                                                                                         \\ \hline
8: \qquad \quad modify the label $l_{tk}$ in L to $k_{t} = argmin(d_{k})$, if    $l_{tk} != argmin(d_{k})$ and $\Delta$ $d_{k}$  $<$ Threshold $\epsilon$ ;                                                                             \\ \hline
9:  \qquad \textbf{end for}                                                                                         \\ \hline
10: \qquad update centroid $c_{k}$ in $C$ to get $C^{'}$ with  \textit{Equation} \ref{EP};                                                                                      \\ \hline
11: \quad \textbf{end while}                                                               \\ \hline    
12: select Top 2 clusters based on the magnitude of the centroid and eliminate the other clusters, update $T$ to get $T^{'}$;
                    \\ \hline  
13: perform crossover ($\bigoplus$ stands for concatenation operation) by picking up ancestors from Top 2 clusters with probability $p_{m}$, \\ to create offspring $o$;
                    \\ \hline 
14: get new population $T^{''}$ $\leftarrow$ $T^{'}$ $\bigcup$ $o$;
            \\ \hline 
15: update the Score set $R$ with new population $T^{''}$ to get $R^{'}$;
            \\ \hline 
16: back to \textbf{line} \textit{5}, repeat until any element in $T$ can have \textit{ASR} = $100.00$;
            \\ \hline 
17: \textbf{return} \textit{T};
            \\ \hline 
\end{tabular}
\end{algorithm*}

To generate effective code snippets, we propose query templates as prompts to ChatGPT, incorporating the important features extracted using attention mechanisms. 
Especially, to better fulfill the requirement of stealthiness, we incorporate the partial codes preceding and succeeding to the inserting locations as context, following the template: \emph{<Context> <Query>< Context>}. And the \emph{<Context>} yields one of the following prompt variations: "Given the partial preceding/succeeding codes as:", "With the partial preceding/following codes provided as:", "In light of the incomplete preceding/following codes as:" and "Taking into account the limited preceding/succeeding codes as:". 
Prior to constructing the queries, we categorize the obtained important features based on their predefined meaning and functionality within the programming language. This allows us to select the appropriate context for each keyword. The categories of the keywords are illustrated in Table \ref{table:category}.
For instance, let's consider the important features "for" (loop-related), "static" (storage type), and "const" (data-type related). Using our predefined templates, we can construct a query such as: "\textit{Please generate a function in C that contains a loop, two structs named $a\_chan$ and $b\_const$, and define a static const char named $a\_rray$ initialized with NULL.}"

To bolster the prevention of vulnerabilities stemming from the introduction of dead code in the adversarial snippet generation process, we append a suffix string indicative of functionality to the obtained significant tokens. For example, if `variable\_a' is utilized as a variable in a conditional statement, `\_condition' is appended to distinguish it before inputting it into the generation module, preventing overlap with the original code. In cases where the keyword does not function as a variable name in the original code, it is nonetheless designated as such in the generated adversarial snippets. Given that the generated code segments are dead code, their operations exhibit no contextual relevance to the surrounding codes. These measures contribute to ultimately defining the generated adversarial snippets as vulnerability-free dead code.

However, the adversarial data generated by ChatGPT often contain more lines than expected, compromising their stealthiness. We aim to improve the concealment by performing in-function insertion with adversarial data. To achieve this, we introduce additional constraints to optimize the prompts. These constraints include the following:
First, we define the request as "\textit{Please generate several lines in C}".
Second, to ensure better integration with the existing code context, we rename the generated structs to be more closely related to the original variable names. For example, we use the optimized query "\textit{Defines two structs $a\_chan$ and $b\_const$ as external structures and rename them as "queue" and "p\_project}".
Third, to reduce the size of the code snippets, we include the condition "\textit{Please generate the codes in dense format}". 
Figure \ref{fig:raw_op} presents an example of the raw data generated by ChatGPT and the optimized data re-generated by ChatGPT after using optimized prompts. 



\subsubsection{Preserved Attack Pool Generation}
%
We have prepared a preserved attack pool that includes meticulously crafted adversarial data generated by ChatGPT. In detail, we generate plenty of samples from ChatGPT. To guarantee the compilability and functionality preservation,  we employ the public program analysis tool (i.e., Comex) to remove these code snippets that are uncompilable or possess data dependency with the original programs and preserve the remaining adversarial code snippets. These adversarial data is specifically designed for studying adversarial attacks and assessing the robustness of machine learning-based software vulnerability detection systems models. 

The preserved attack pool contains all the samples generated based on five categories of important features as discussed in subsection \ref{chatgpt}. These important features are  data type, control statement, storage classes, input-output, and miscellaneous. These samples will be used as seed input for the fuzzy genetic algorithm to optimize the attack strategy. In other words, by using the preserved attack pool as a source of seed samples, fuzzy genetic algorithms can leverage the knowledge and characteristics of previously crafted adversarial examples. These seed data will provide desirable properties, such as effective attack strategies or high success rates against machine learning models.

\subsection{Adversarial Learning}
\label{evasion_attack}
This subsection will discuss Step 2, which involves adversarial learning, including seed data selection and evasion attacks. 

\subsubsection{Seed Data Selection using FGA}

To discover the optimal combination of preserved templates and enhance the success rate of evasion attacks, we employ optimized Fuzzy Genetic Algorithm (FGA) \cite{wu2020multitasking} (Algorithm 1). 
The FGA method employs a fuzzy clustering approach to ensure that all reserved members in the population have an opportunity to pass on to the next generation. Additionally, a fuzzy selection method is utilized to mitigate the drawbacks of a greedy strategy. The novel genetic algorithm consists of four major steps: initialization, clustering, selection, and crossover.
Algorithm \ref{table:FAG} provides detailed information regarding the seed data generation.

\textit{Initialization.} 
The genetic algorithm begins by randomizing the initial population, which serves as the first step in our proposed optimization algorithm. Each sample within the randomly generated population is filled with the predefined code snippets based on the templates. The population size remains constant throughout the method. Additionally, we initialize the score set of the population by calculating the fitness scores for each member.
Furthermore, we randomly sample a centroid set from a uniform distribution within the range of $[0, 1)$, denoted as $C = (c_{1}, c_{2}, ..., c_{k})$, where $k$ represents the number of clusters.

\textit{Fitness Function.}
The design of the fitness function is a crucial step in genetic algorithms as it greatly affects the inheritance and success rate of the algorithm. In our proposed approach, we incorporate the attack success rate and the length of inserted code snippets into the fitness function. Intuitively, an effective adversarial sample should have a higher attack success rate and a lower length of inserted code snippets. Such samples are more likely to be selected as mating candidates or as the desired outcome. Based on this idea, the calculation of the fitness score for each member of the population is formulated as follows:
\begin{equation}\label{ES}  
Score(s_{j}) = ASR(D_{i} \otimes s_{j}) - \lambda * len(s_{j}) 
\end{equation}
where $ASR(D_{i} \otimes s_{j})$ is the averaged attack success rate of entities generated by inserting the statement snippets into test vulnerable programs and $len(s_{j})$ is the number of lines of the sample in the population. $\lambda$ controls the significance of the lengths of code snippets.

\begin{figure*}[!tb]\centering
	\subfloat[Vulnerable and non-vulnerable (normal) samples \label{Fig.V1}]
	{\includegraphics[width=0.5\textwidth]{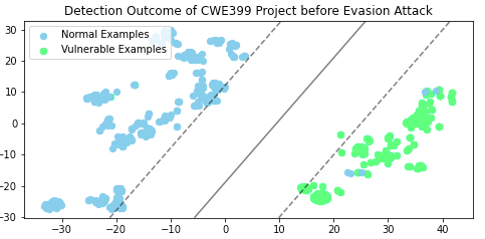}}
	\subfloat[Movement of adversarial samples  \label{Fig.V2}]
	{\includegraphics[width=0.5\textwidth]{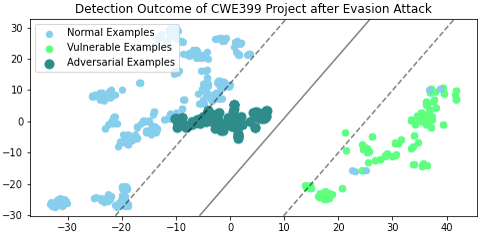}}
	\caption{Visualization of Evasion Attack using EaTVul. This figure shows the distribution of vulnerable and non-vulnerable samples, as well as the adversarial samples, based on t-SNE. 
	}
	\label{Fig.V}
\end{figure*}

\textit{Fuzzy clustering approach.} Fuzzy clustering \cite{miyamoto2008algorithms} is a type of clustering algorithm that assigns each data point to multiple clusters with corresponding probabilities instead of a single cluster. In our work, we employ fuzzy clustering for further selection of the mating pool, aiming to avoid sub-optimal combinations.
Compared to conventional clustering methods, where each sample in the population is assigned to a single cluster, denoted as $y = (y_{1}, y_{2}, y_{3}, ..., y_{n})$, the fuzzy clustering algorithm defines rules to separate the data into clusters $C = (c_{1}, c_{2}, c_{3}, ..., c_{m})$ in a way that minimizes the overall loss. The loss is defined by the following equation:
\begin{equation}\label{EL1}  
argmin_{C} =  \sum^{n}_{j=1} \sum^{c}_{k=1} w^{\alpha}_{jk} |score(s_{j}) - c_{k} |
\end{equation}
where the fuzziness is limited by the factor $\alpha$. The partition matrix is denoted by $W = w_{ij}$, which indicates the probability that element $y_{i}$ belongs to $C_{j}$. And $w_{ij}$ is calculated as:
\begin{equation}\label{EL2}  
w_{jk} = \frac{1}{\sum^{c}_{k=1}(\frac{s_{j} - c{j}}{s_{j} - c_{k}})^{^{\frac{2}{C-1}}}}
\end{equation}
where $C$ is the number of clusters.

\textit{Selection.}
We select the best two clusters based on the magnitude of the centroid, which is also known as the fitness score. Our intuition is that the mating candidate with a higher fitness score has a higher probability of creating high-quality and effective sub-generations. However, it is worth noting that all members in the population have a chance to participate in the reproduction process, but with a lower probability.

\textit{Crossover.} The operator randomly samples two chromosomes as parents from the selected clusters, ensuring that replicated code snippet combinations are discarded. The probability of a chromosome being selected as a parent is calculated as follows:
\begin{equation}\label{EP}  
p_{i} = \frac{e^{f_{i}}}{\sum^{n}_{k=1} e^{f_{i}}}
\end{equation}
where $f_{i} = w^{\alpha}_{ik} |score(s_{i}) - c_{k} |$ and $c_{k}$ denotes the centroid of the selected clusters.


\subsubsection{Evasion Attack}

When launching an attack on a vulnerable test case, the FGA randomly chooses a seed sample. It then produces an optimized adversarial data snippet, which is added to the vulnerable test case, which mean we add the code snippet into the vulnerable samples before feed it to the prediction model. The objective is to modify the test case in such a way that it can evade detection by the machine learning-based software vulnerability detection system.
It should be emphasized that EaTVul is specifically designed for evasion attacks targeting vulnerable test cases. 
The consolidation of all the keywords from Figure \ref{Fig.C1} into a single function is illustrated in Figure \ref{fig:raw_op}, as shown in Figure \ref{Fig.C2}, while maintaining the core logic. The code depicted in Figure \ref{Fig.C2} will be incorporated into the non-vulnerable sample to initiate the attack.

In addition, it is worth noting that, for the input and output of each step in the proposed EaTVul, the generated results are consistent with expectations at each stage. Specifically, given the collected training data, we employ SVM to identify the most important non-vulnerable samples. SVM consistently outputs the same support vectors in this step. Subsequently, the attention mechanism is applied to identify the most important features given these crucial samples. The attention mechanism consistently produces the same important features or tokens in this step. Afterwards, we set up rules for ChatGPT to generate adversarial data. The temperature for ChatGPT is set to 0, ensuring reproducibility in the results at this step. All generated adversarial data is added to the preserved attack pool, maintaining consistency with the previous step. Next, a fuzzy genetic algorithm is applied to create the best attack strategy, consistently producing the best combination based on the samples from the same preserved attack pool. Therefore, this process is reproducible as it is automatically completed based on FGA. Finally, we launch an evasion attack based on the adversarial data created by FGA. Since each process can be reproduced, we can confirm that the generated results are consistent and reproducible, given the same training data.


\subsection{Visualization of EaTVul evasion attack}
In this subsection, we present the visualization of the data distribution using t-SNE \cite{van2008visualizing} given vulnerable and non-vulnerable samples. 
%
%
Figure \ref{Fig.V} showcases the application of t-SNE for visualizing evasion attacks, specifically focusing on the vulnerable and non-vulnerable features. The visualization presented in Figure \ref{Fig.V1} reveals interesting insights regarding the separability of vulnerable and non-vulnerable samples based on the decision boundary, with only a few vulnerable samples overlapping.
By leveraging t-SNE, we can gain a deeper understanding of the effectiveness of evasion attacks and the distinguishability of vulnerable and non-vulnerable instances. The clear separation observed in the visualization indicates that the decision boundary is generally successful in classifying samples as vulnerable or non-vulnerable.


However, it is worth noting that despite the initial separation observed in Figure \ref{Fig.V1}, certain vulnerable samples may actually lie within the region classified as non-vulnerable. This discrepancy can be attributed to the presence of precisely crafted adversarial data. By strategically inserting adversarial perturbations into the vulnerable samples, we can observe a shift in their positions towards the non-vulnerable side, as depicted in the Figure\ref{Fig.V2}. This demonstrates the susceptibility of vulnerable samples to manipulation and highlights the need for robust defense mechanisms capable of mitigating such adversarial attacks.

\section{Experimental Setup}
We will fundamentally evaluate the proposed EaTVul by answering the following research questions (RQ):
\begin{itemize} 
    \item How effective is fuzzy genetic algorithm in selecting the seed adversarial data compared with randomization? 
    \item How effective is EaTVul based on adversarial data generated by ChatGPT originally and after optimization? 
    To answer this question, we conducted experiments by presenting the results based on the scenario using the original adversarial data directly and optimized data. 
    \item How effective is EaTVul with recently developed machine learning-based software vulnerability detection systems? 
    \item How effective is EaTVul when compared with state-of-the-art large language models (LLM) and other machine learning tools that using BiLSTM for software vulnerability detection? To answer this question, we compare EaTVul with four recent large program generation models/algorithms: CodeBERT \cite{hin2022linevd}, CodeGen-2B \cite{chan2023transformer}, Poster-Lin \cite{lin2017poster}, and MDVD \cite{lin2020deep}.  
    \item How EaTVul behaves/performs regarding obfus-
cation/diversification methods? To answer this question, We conducted comparative experiments on two state-of-the-art target models, Asteria \cite{yang2021asteria} and LineVul \cite{fu2022linevul}, which leverage different program representations.
    \item How effective is EaTVul when generalized to other
programming languages? To demonstrate the generalizability of our proposed method, we extend the evaluation to two state-of-the-art vulnerability detection models for Java programs, namely FUNDED \cite{wang2020combining} and VDet \cite{mamede2022transformer}. 
\end{itemize}

\subsection{Datasets}
\label{dataset_setting}
In this study, we consider using multiple datasets in the C/C++ programming language including the real-world Asterisk project and OpenSSL project. These two projects include multiple types of vulnerabilities and are widely used in the community \cite{li2021pdgraph,li2016vulpecker}.
For the scenario of a single type of vulnerability, we reuse the datasets from \cite{li2018vuldeepecker, mirsky2023vulchecker}, which are  CWE119 and CWE399 datasets. CWE119 is a buffer error dataset, CWE399 is a resource management errors dataset, and CWE416 is a use-after-free vulnerability dataset. 
Considering the SARD dataset  \cite{feng2020efficient}, \cite{yang2021asteria}, \cite{fu2022linevul} has been widely used in the area of software vulnerability detection, experiments based on this dataset has been conducted as well. To demonstrate the generalizability of our proposed method, we systematically assess its performance across Java code samples sourced from the National Vulnerability Database (NVD) and open-source projects on GitHub, specifically targeting those classified within the top 5 to top 30 most perilous categories as defined in the Common Weakness Enumeration (CWE) \cite{wang2020combining}.

Table \ref{dataset} presents the dataset information used in this paper. The first column shows the name of the dataset, the second column shows the number of vulnerable samples, followed by the number of non-vulnerable samples. 
In the experimental settings of training the surrogate model, we split the dataset into training (70\%), evaluation (15\%), and test data (15\%). 
All the vulnerable test cases come from the test data. All the vulnerable test cases have been tested and confirmed that they are classified as vulnerable by the target model before launching adversarial attacks. 
The Abstract Syntax Trees (AST) feature has been considered in this paper since all the baselines use AST in their study.

\begin{table}[!tb]
\centering
\caption{Dataset Information.}
\label{dataset}
\begin{tabular}{|l|c|c|c|}
\hline
Data     & \multicolumn{1}{l|}{\#Vulnerable} & \multicolumn{1}{l|}{\#non-vulnerable} & \multicolumn{1}{l|}{\# Total} \\ \hline
Asterisk & 102                                & 1541                                 & 1553                         \\ \hline
OpenSSL  & 157                               & 788                                   & 945                           \\ \hline
CWE119   & 5442                             & 6966                                  & 12388                         \\ \hline
CWE399   & 1232                              & 1288                                  & 2520                          \\ \hline
SARD   & 64788                              & 131792                                  & 196580                          \\ \hline
CWE416   & 459                              & 1834                                  & 2293                          \\ \hline
JAVA   & 14756                              & 14756                                  & 29512                          \\ \hline
\end{tabular}
\end{table}

\subsection{Evaluation Metrics}
This work only considers the attack success rate, which is defined as follows:
\begin{equation}\label{PM}  
\frac{\# \: bypass}{\# \: total \: test \: cases} \times 100\%
\end{equation}

In this study, the success rate typically refers to the percentage of attempts or instances where an adversarial attack is successful in bypassing the target model's (i.e., the vulnerable test cases misclassified as non-vulnerable). 
The success rate quantifies how often the attack is able to generate adversarial examples that are misclassified by the target model. It represents the proportion of crafted adversarial samples that successfully fool the model into making incorrect predictions or classifications. 
A higher success rate indicates a stronger attack, as it demonstrates the ability to consistently manipulate the  target model's predictions. 

Considering some baselines only report the top@k precision, we have also included experimental results based on top@k in our report. For instance, a success rate of 90\% in Top@10 means that the attack was successful in 9 out of 10 cases when considering only the top 10 most vulnerable instances.
It is important to note that reporting the top@k precision gives us insights into the attack`s effectiveness in specific scenarios where only a limited number of vulnerable instances are of interest. 

In this study, the snippet size \cite{jain2023code} is an additional factor considered to improve the success rate of the attack in specific circumstances. The snippet size refers to the number of individual pieces of adversarial data that are chosen and combined to create a single adversarial sample used for launching the attack. For instance, when the snippet size is 1, it means that only one piece of adversarial data is selected and added to the vulnerable sample. 
On the other hand, when the snippet size is 2, two pieces of adversarial data are chosen and combined together to form a single adversarial sample. This allows for more complex modifications and combinations of multiple features or aspects in the attack process.

In addition, F1-Score \cite{liu2019cyber} Without Adversarial Samples has been considered in this study as well. As stated above, the adversary's goal is to degrade the performance of the input code sequence with malicious code snippets by imperceptibly modifying the original source code, while maintaining the detection capability on the samples without adversarial code snippets. Therefore, we employ the F1-score without adversarial attack to represent the normal performance in the downstream vulnerability detection task.


\begin{table}[!tb]
\centering
\caption{Experimental results of attack success rate (\%) based on different top@k metrics of fuzzy genetic algorithm and randomization selection.}
\label{ran_vs_genetic}
\begin{tabular}{|l|c|c|c|c|}
\hline
Strategy      & \multicolumn{1}{l|}{Top@5} & \multicolumn{1}{l|}{Top@10} & \multicolumn{1}{l|}{Top@15} & \multicolumn{1}{l|}{Top@20} \\ \hline
Randomizat. & 0.533                      & 0.475                       & 0.464                       & 0.454                       \\ \hline
Genetic       & 1.000                      & 0.925                       & 0.858                       & 0.798                       \\ \hline
\end{tabular}
\end{table}

\subsection{Baselines}
In this work, we choose the recently developed systems as the target models to show the effectiveness of the proposed EaTVul scheme. The baselines include AST-EVD \cite{feng2020efficient}, Asteria \cite{yang2021asteria}, LineVul \cite{fu2022linevul}, Poster-Lin \cite{lin2017poster}, MDVD \cite{lin2020deep}, CodeBERT \cite{hin2022linevd}, and CodeGen-2B \cite{chan2023transformer}.  
POSTER-Lin employs a customized bi-directional LSTM network for function-level vulnerability discovery using the AST. MDVD effectively detects function-level vulnerabilities by combining heterogeneous data sources to extract transferable vulnerability information and utilizing LSTM cells to learn unified representations of vulnerable source code patterns. AST-EVD avoids data loss by using the pack-padded method on the Bi-GRU network, achieving precise function-level vulnerability detection. Asteria utilizes the Tree-LSTM and code pairing to identify vulnerabilities in defective programs. LineVul proposes a transformer-based method with an attention mechanism for comprehensive line-level vulnerability detection, utilizing context information effectively. These approaches enhance vulnerability assessment using different techniques and data sources. CodeBERT is a pre-trained programming language model that encompasses multiple programming languages. The CodeGen-2B transformer decoder model is trained on a variety of programming languages, including C, C++, Go, Java, JavaScript, and Python, as well as natural language

Previous research treats obfuscation transformations to programs as adversarial perturbations that can affect a downstream ML/DL model like a malware detector or a program summarizer. Since our proposed attack schema operates on the natural code channel, we compare it with two state-of-the-art obfuscation techniques for a fair comparison, as they also focus on the source code: Differentiable Generator \cite{Srikant2021GeneratingAC}, which uses first-order gradient information to discover the optimal choice of sites in a program for applying perturbations and the specific perturbations to apply on the selected sites; Milo \cite{song2023milo}, which supports five obfuscation methods that alter the semantic and syntactic features of the programs. Specifically, these methods fall into two types of transformation: replacing transformation and insert transformation.

\begin{table}[!tb]
\centering
\caption{Experimental results of the attack success rate (\%) based on adversarial data generated by ChatGPT, both originally and after optimization.}
\label{table:data}
\begin{tabular}{|l|l|c|c|}
\hline
Models                   & Data    & \multicolumn{1}{l|}{Original data} & \multicolumn{1}{l|}{Optimized data} \\ \hline
\multirow{3}{*}{AST-EVD} & SARD    & 0.879                              & 1.000                               \\ \cline{2-4} 
                         & OpenSSL & 0.825                              & 0.943                               \\ \cline{2-4} 
                         & CWE399  & 0.889                              & 0.975                               \\ \hline
\multirow{3}{*}{Asteria} & SARD    & 0.902                              & 1.000                               \\ \cline{2-4} 
                         & OpenSSL & 0.854                              & 0.980                               \\ \cline{2-4} 
                         & CWE399  & 0.922                              & 1.000                               \\ \hline
\multirow{3}{*}{LineVul} & SARD    & 0.845                              & 0.983                               \\ \cline{2-4} 
                         & OpenSSL & 0.856                              & 1.000                               \\ \cline{2-4} 
                         & CWE399  & 0.837                              & 0.956                               \\ \hline
\end{tabular}
\end{table}

\subsection{Results and Discussion}
We structure our evaluation by stepping through
each of our three research questions (RQ1-6).

\textit{\textbf{RQ1: How effective is fuzzy genetic algorithm in selecting the seed adversarial data compared with randomization?}}
In order to show that fuzzy genetic algorithm outperforms randomization selection, we conducted experiments on the four datasets (i.e., CWE119, CWE399, Asterisk, and OpenSSL) and report the average results. 
%
Table \ref{ran_vs_genetic} presents the experimental results of fuzzy genetic algorithm and randomization selection. The provided results show the performance of two strategies, Randomization and Genetic, based on different top@k metrics (Top@5, Top@10, Top@15, and Top@20).

For the Randomization strategy, the success rates gradually decrease as we consider a larger number of vulnerable instances. At Top@5, the success rate is 53.3\%, indicating that 53.3\% of the top 5 vulnerable cases were successfully attacked. As we expand the evaluation to the top 10, 15, and 20 vulnerable instances, the success rates decline further to 47.5\%, 46.4\%, and 45.4\%, respectively. This suggests that the Randomization strategy performs relatively less effectively as we consider a larger pool of vulnerable cases.

On the other hand, EaTVul with the Genetic strategy demonstrates high success rates across all top@k metrics. At Top@5, the success rate is a perfect 100\%, indicating that all of the top 5 vulnerable instances were successfully attacked. Even as we consider a larger number of vulnerable instances, EaTVul with the Genetic strategy maintains high success rates. At Top@10, the success rate is 92.5\%, which means that 92.5\% of the top 10 vulnerable cases were successfully attacked. 
Although there is a slight decrease in the attack success rates for top@15 and top@20, with 85.8\% and 79.8\% respectively, the results still demonstrate the effectiveness of EaTVul with the Genetic strategy when compared to the Randomization strategy.

\begin{table}[!tb]
\centering
\caption{Experimental results of attack success rate (\%) against AST-EVD, Asteria, and LineVul. Sni.: Snippet Size.}
\label{table:three_target_model}
\begin{tabular}{|l|l|c|c|c|c|}
\hline
{\color[HTML]{000000} \begin{tabular}[c]{@{}l@{}}Target\\ Model\end{tabular}} & {\color[HTML]{000000} Dataset} & \multicolumn{1}{l|}{Sni.-1} & \multicolumn{1}{l|}{Sni.-2} & \multicolumn{1}{l|}{Sni.-3} & \multicolumn{1}{l|}{Sni.-4} \\ \hline
                                                                              & Asterisk                       & 0.480                       & 0.750                       & 0.840                       & 1.000                       \\ \cline{2-6} 
                                                                              & OpenSSL                        & 0.520                       & 0.790                       & 0.840                       & 1.000                       \\ \cline{2-6} 
                                                                              & CWE119                         & 0.510                       & 0.630                       & 0.780                       & 0.860                       \\ \cline{2-6} 
                                                                              & CWE399                         & 0.570                       & 0.830                       & 0.920                       & 1.000                       \\ \cline{2-6} 
\multirow{-5}{*}{\begin{tabular}[c]{@{}l@{}}AST-\\ EVD\end{tabular}}          & SARD                           & 0.460                       & 0.650                       & 0.840                       & 0.920                       \\ \hline
                                                                              & Asterisk                       & 0.620                       & 0.740                       & 0.880                       & 1.000                       \\ \cline{2-6} 
                                                                              & OpenSSL                        & 0.650                       & 0.710                       & 0.930                       & 1.000                       \\ \cline{2-6} 
                                                                              & CWE119                         & 0.670                       & 0.860                       & 1.000                       & 1.000                       \\ \cline{2-6} 
                                                                              & CWE399                         & 0.760                       & 0.830                       & 0.960                       & 1.000                       \\ \cline{2-6} 
\multirow{-5}{*}{Asteria}                                                     & SARD                           & 0.640                       & 0.840                       & 0.950                       & 1.000                       \\ \hline
                                                                              & Asterisk                       & 0.340                       & 0.730                       & 0.940                       & 1.000                       \\ \cline{2-6} 
                                                                              & OpenSSL                        & 0.340                       & 0.910                       & 0.970                       & 1.000                       \\ \cline{2-6} 
                                                                              & CWE119                         & 0.370                       & 0.560                       & 0.910                       & 1.000                       \\ \cline{2-6} 
                                                                              & CWE399                         & 0.430                       & 0.740                       & 0.830                       & 0.930                       \\ \cline{2-6} 
\multirow{-5}{*}{LineVul}                                                     & SARD                           & 0.450                       & 0.670                       & 0.870                       & 1.000                       \\ \hline
\end{tabular}
\end{table}

\begin{table*}[!tb]
\centering
\caption{Experimental results of attack success rate (\%) against Poster-Lin, MDVD, CodeBERT, and CodeGen.}
\label{table:BiLSTM}
\begin{tabular}{|l|l|cccc|cccc|}
\hline
\multicolumn{1}{|c|}{\multirow{2}{*}{\begin{tabular}[c]{@{}c@{}}Target\\    Model\end{tabular}}} & \multicolumn{1}{c|}{\multirow{2}{*}{Data}} & \multicolumn{4}{c|}{Snippet Size = 2}                                                             & \multicolumn{4}{c|}{Snippet Size =3}                                                              \\ \cline{3-10} 
\multicolumn{1}{|c|}{}                                                                           & \multicolumn{1}{c|}{}                      & \multicolumn{1}{c|}{Top@5} & \multicolumn{1}{c|}{Top@ 10} & \multicolumn{1}{c|}{Top@15} & Top@ 20 & \multicolumn{1}{c|}{Top@ 5} & \multicolumn{1}{c|}{Top@ 10} & \multicolumn{1}{c|}{Top@15} & Top@20 \\ \hline
\multirow{4}{*}{Poster-Lin}                                                                      & Asterisk                                   & \multicolumn{1}{c|}{0.900} & \multicolumn{1}{c|}{0.867}   & \multicolumn{1}{c|}{0.756}  & 0.758   & \multicolumn{1}{c|}{1.000}  & \multicolumn{1}{c|}{1.000}   & \multicolumn{1}{c|}{0.878}  & 0.858  \\ \cline{2-10} 
                                                                                                 & OpenSSL                                    & \multicolumn{1}{c|}{1.000} & \multicolumn{1}{c|}{1.000}   & \multicolumn{1}{c|}{0.745}  & 0.717   & \multicolumn{1}{c|}{1.000}  & \multicolumn{1}{c|}{1.000}   & \multicolumn{1}{c|}{0.911}  & 0.875  \\ \cline{2-10} 
                                                                                                 & CWE119                                     & \multicolumn{1}{c|}{0.933} & \multicolumn{1}{c|}{0.934}   & \multicolumn{1}{c|}{0.899}  & 0.867   & \multicolumn{1}{c|}{1.000}  & \multicolumn{1}{c|}{0.967}   & \multicolumn{1}{c|}{0.956}  & 0.925  \\ \cline{2-10} 
                                                                                                 & CWE399                                     & \multicolumn{1}{c|}{1.000} & \multicolumn{1}{c|}{0.833}   & \multicolumn{1}{c|}{0.823}  & 0.767   & \multicolumn{1}{c|}{1.000}  & \multicolumn{1}{c|}{1.000}   & \multicolumn{1}{c|}{1.000}  & 0.917  \\ \hline
\multirow{4}{*}{MDVD}                                                                            & Asterisk                                   & \multicolumn{1}{c|}{1.000} & \multicolumn{1}{c|}{0.867}   & \multicolumn{1}{c|}{0.844}  & 0.784   & \multicolumn{1}{c|}{1.000}  & \multicolumn{1}{c|}{1.000}   & \multicolumn{1}{c|}{1.000}  & 0.975  \\ \cline{2-10} 
                                                                                                 & OpenSSL                                    & \multicolumn{1}{c|}{1.000} & \multicolumn{1}{c|}{1.000}   & \multicolumn{1}{c|}{0.845}  & 0.817   & \multicolumn{1}{c|}{1.000}  & \multicolumn{1}{c|}{1.000}   & \multicolumn{1}{c|}{1.000}  & 1.000  \\ \cline{2-10} 
                                                                                                 & CWE119                                     & \multicolumn{1}{c|}{1.000} & \multicolumn{1}{c|}{0.933}   & \multicolumn{1}{c|}{0.877}  & 0.825   & \multicolumn{1}{c|}{1.000}  & \multicolumn{1}{c|}{1.000}   & \multicolumn{1}{c|}{0.978}  & 0.958  \\ \cline{2-10} 
                                                                                                 & CWE399                                     & \multicolumn{1}{c|}{1.000} & \multicolumn{1}{c|}{0.899}   & \multicolumn{1}{c|}{0.867}  & 0.767   & \multicolumn{1}{c|}{1.000}  & \multicolumn{1}{c|}{1.000}   & \multicolumn{1}{c|}{0.967}  & 0.950  \\ \hline
\multirow{4}{*}{CodeBERT}                                                                        & Asterisk                                   & \multicolumn{1}{c|}{0.900} & \multicolumn{1}{c|}{0.850}   & \multicolumn{1}{c|}{0.803}  & 0.740   & \multicolumn{1}{c|}{0.900}  & \multicolumn{1}{c|}{0.885}   & \multicolumn{1}{c|}{0.845}  & 0.832  \\ \cline{2-10} 
                                                                                                 & OpenSSL                                    & \multicolumn{1}{c|}{0.800} & \multicolumn{1}{c|}{0.800}   & \multicolumn{1}{c|}{0.768}  & 0.735   & \multicolumn{1}{c|}{0.900}  & \multicolumn{1}{c|}{0.864}   & \multicolumn{1}{c|}{0.858}  & 0.835  \\ \cline{2-10} 
                                                                                                 & CWE119                                     & \multicolumn{1}{c|}{0.900} & \multicolumn{1}{c|}{0.840}   & \multicolumn{1}{c|}{0.834}  & 0.785   & \multicolumn{1}{c|}{1.000}  & \multicolumn{1}{c|}{0.935}   & \multicolumn{1}{c|}{0.911}  & 0.865  \\ \cline{2-10} 
                                                                                                 & CWE399                                     & \multicolumn{1}{c|}{0.900} & \multicolumn{1}{c|}{0.825}   & \multicolumn{1}{c|}{0.786}  & 0.776   & \multicolumn{1}{c|}{1.000}  & \multicolumn{1}{c|}{0.920}   & \multicolumn{1}{c|}{0.878}  & 0.858  \\ \hline
\multirow{4}{*}{CodeGen}                                                                         & Asterisk                                   & \multicolumn{1}{c|}{0.900} & \multicolumn{1}{c|}{0.867}   & \multicolumn{1}{c|}{0.844}  & 0.784   & \multicolumn{1}{c|}{0.933}  & \multicolumn{1}{c|}{0.925}   & \multicolumn{1}{c|}{0.899}  & 0.867  \\ \cline{2-10} 
                                                                                                 & OpenSSL                                    & \multicolumn{1}{c|}{0.900} & \multicolumn{1}{c|}{1.000}   & \multicolumn{1}{c|}{0.845}  & 0.817   & \multicolumn{1}{c|}{0.956}  & \multicolumn{1}{c|}{0.911}   & \multicolumn{1}{c|}{0.875}  & 0.845  \\ \cline{2-10} 
                                                                                                 & CWE119                                     & \multicolumn{1}{c|}{1.000} & \multicolumn{1}{c|}{0.933}   & \multicolumn{1}{c|}{0.877}  & 0.825   & \multicolumn{1}{c|}{1.000}  & \multicolumn{1}{c|}{1.000}   & \multicolumn{1}{c|}{0.928}  & 0.875  \\ \cline{2-10} 
                                                                                                 & CWE399                                     & \multicolumn{1}{c|}{1.000} & \multicolumn{1}{c|}{0.899}   & \multicolumn{1}{c|}{0.867}  & 0.767   & \multicolumn{1}{c|}{1.000}  & \multicolumn{1}{c|}{0.950}   & \multicolumn{1}{c|}{0.917}  & 0.880  \\ \hline
\end{tabular}
\end{table*}

In summary, the results indicate that EaTVul with the Genetic strategy outperforms the Randomization strategy across all top@k metrics. It achieves higher success rates and demonstrates greater effectiveness in attacking a larger number of vulnerable instances. These findings highlight the potential of EaTVul with the Genetic strategy as a robust approach for adversarial attacks in the given context.

\textit{\textbf{RQ2: How effective is EaTVul based on adversarial data generated by ChatGPT originally and after optimization?}}
To answer this question, we conducted experiments using the adversarial data generated from ChatGPT directly and the optimized data by ChatGPT. We only report the experimental results based on the OpenSLL, CWE399, and SARD datasets since the experimental results are similar for the Asterisk and CWE119 datasets.

Table \ref{table:data} presents the performance of attacking different models of EaTVul on various datasets using both the original data generated from ChatGPT and the re-generated optimized data by ChatGPT. The models evaluated include AST-EVD, Asteria, and LineVul, while the datasets examined are SARD dataset, OpenSSL dataset, and CWE399 dataset.

For attacking AST-EVD, EaTVul achieves an attack success rate of 0.879 on the SARD dataset when using the original data. However, this attack success rate significantly improved to a perfect 1.000 (i.e., 100\%) when the optimized data was utilized. Similarly, for the OpenSSL dataset, the success rate increased from 0.825 (original data) to 0.943 (optimized data). In the case of the CWE399 dataset, the success rate improved from 0.889 (original data) to 0.975 (optimized data). These results clearly indicate that optimizing the data has a positive impact on the performance of the models. Overall, the findings demonstrate the effectiveness of data optimization in enhancing the attack success rates against the AST-EVD model across different datasets.

Similarly, for the Asteria model, EaTVul demonstrates improved attack success rates with data optimization. On the SARD dataset, the original data achieves a success rate of 0.902, while the optimized data achieves a perfect 1.000. Similarly, on the OpenSSL dataset, the attack success rate increases from 0.854 with the original data to 0.980 with the optimized data. The CWE399 dataset also shows significant improvement, with the original data achieving a success rate of 0.922, which rises to a perfect 1.000 (i.e., 100\%) with the optimized data. These results highlight the effectiveness of data optimization in enhancing the attack success rates of EaTVul against Asteria model. Moreover, for the LineVul model, the attack success rate of EaTVul on the SARD dataset increases from 0.845 with the original data to 0.983 with the optimized data. On the OpenSSL dataset, the success rate improves from 0.856 to a perfect 1.000. The CWE399 dataset also shows improvement, with the success rate increasing from 0.837 to 0.956. These results emphasize the effectiveness of data optimization in enhancing the attack success rates of EaTVul against the LineVul model across multiple datasets, leading to improved attack performance in adversarial learning of vulnerability detection.

Overall, the results indicate that the optimization of data improves the performance of the models across all datasets. The higher attack success rates achieved with the optimized data demonstrate the effectiveness of the optimization process in enhancing the EaTVul models' attack capabilities.
This in turn emphasized the important of our proposed optimization process.

\textit{\textbf{RQ3: How effective of EaTVul towards recently developed machine learning-based vulnerability detection systems?}}
To answer this question, we have conducted experiments with baselines AST-EVD \cite{feng2020efficient}, Asteria \cite{yang2021asteria}, LineVul \cite{fu2022linevul}.

Table \ref{table:three_target_model} presents the results of experiments conducted on different target models using various datasets and snippet sizes. The target models include AST-EVD, Asteria, and LineVul. The datasets used in the experiments are Asterisk, OpenSSL, CWE119, and CWE399. It is worth noting that the dataset SARD has been considered in this experiment given that all three baselines have used SARD dataset in their study. For each target model and dataset combination, the table provides the success rates of the attack for different snippet sizes, ranging from 1 to 4. A higher success rate indicates a greater effectiveness of the attack in deceiving the target model.

Generally speaking, the success rates of the EaTVul attack strategy on the different target models vary depending on the dataset and snippet size. For example, for the Asterisk dataset of AST-EVD, Asteria, and LineVul models, the attack success rates of EaTVul range from 48.0\%, 62.0\% and 34.0\% for snippet size -1 to a perfect 100.0\% for snippet size -4 for all the target models. Similar results happens to scenarios except for AST-EVD on CWE-119 and SARD datasets, and LineVul on the CWE399 dataset. 
Looking at the overall trends in the table, it can be observed that higher snippet sizes generally lead to higher success rates. This suggests that incorporating multiple modifications or combinations of adversarial data enhances the effectiveness of the EaTVul. 
For some combinations, the success rates reach 100\%, indicating a complete success in evasion attack. 

Overall, the table provides valuable insights into the performance of the EaTVul attack strategy across various target models, datasets, and snippet sizes. It helps in understanding the effectiveness of the proposed EaTVul attack in different scenarios and can guide researchers and practitioners in optimizing their strategies for adversarial attacks.

\textit{\textbf{RQ4: How effective is EaTVul when compared with State-of-the-art large language models (LLM) and other machine learning tools that using BiLSTM for software vulnerability detection?}}
To answer this question, we conducted experiment based on the datasets of CWE119, CWE399, Asterisk and OpenSSL. We report the top-k precision to keep the same performance measure as used in the baselines.

Table \ref{table:BiLSTM} presents the results of an attack strategy on different target models using specific datasets and two different snippet sizes: Snippet Size = 2 and Snippet Size = 3. The performance is evaluated based on the top@5, top@10, top@15, and top@20 metrics.

For the target model "Poster-Lin," with the Asterisk dataset, the attack success rates of EaTVul range from 90.0\% to 75.8\% for snippet size 2, and from 100.0\% to 85.8\% for snippet size 3. When applied to the OpenSSL dataset, the attack success rates range from 100.0\% to 71.7\% for snippet size 2 and from 100.0\% to 87.5\% for snippet size 3. Similarly, for the CWE119 dataset, the attack success rates range from 93.3\% to 86.7\% for snippet size 2, and from 100.0\% to 92.5\% for snippet size 3. The CWE399 dataset yields success rates of 100.0\% to 76.7\% for snippet size 2 and 100.0\% to 91.7\% for snippet size 3. In the case of the "MDVD" target model, using the Asterisk dataset, the attack success rates of EaTVul range from 100.0\% to 78.4\% for snippet size 2 and from 100.0\% to 97.5\% for snippet size 3. With the OpenSSL dataset, the attack success rates remain consistently high at 100.0\% across all metrics and snippet sizes. The CWE119 dataset produces success rates ranging from 100.0\% to 82.5\% for snippet size 2, and from 100.0\% to 95.8\% for snippet size 3. Finally, the CWE399 dataset demonstrates high success rates ranging from 100.0\% to 76.7\% for a snippet size of 2, and from 100.0\% to 95.0\% for a snippet size of 3. Despite the variation in attack success rates, the attacks of EaTVul consistently maintain a high level of effectiveness.

For the target model "CodeBERT model", there's a consistent trend indicating improved performance with larger snippet sizes. For instance, in the Asterisk dataset, the success rates for top@5, top@10, top@15, and top@20 increase from 90.0\%, 85.0\%, 80.3\%, and 74.0\%, respectively, for Snippet Size = 2, to 90.0\%, 88.5\%, 84.5\%, and 83.2\%, respectively, for Snippet Size = 3. This trend persists across different datasets like OpenSSL, CWE119, and CWE399, where larger snippet sizes consistently lead to higher success rates. For instance, in the CWE119 dataset, the success rates for top@5, top@10, top@15, and top@20 improve from 90.0\%, 84.0\%, 83.4\%, and 78.5\%, respectively, for Snippet Size = 2, to 100.0\% across all metrics for Snippet Size = 3. The "CodeGen" model exhibits similar trends, with varying success rates across datasets and snippet sizes. 

\begin{table}[!t]
\centering
\caption{Experimental results of ASR and F1-Score against Asteria and LineVul with obfuscation techniques. ASR:Attack Success Rate (\%); F1-Score: F1-Score of Target Model Without Adversarial Samples (\%).}
\begin{tabular}{|l|l|l|c|c|}
\hline
Dataset                 & \begin{tabular}[c]{@{}l@{}}Target \\ Model\end{tabular} & \begin{tabular}[c]{@{}l@{}}Attack \\ Model\end{tabular}              & \multicolumn{1}{l|}{ASR} & \multicolumn{1}{l|}{F1-Score} \\ \hline
\multirow{7}{*}{CWE119} & \multirow{3}{*}{Asteria}                                & \begin{tabular}[c]{@{}l@{}}Differentiable \\ Obfuscator\end{tabular} & 66.40                    & \multirow{3}{*}{81.45}        \\ \cline{3-4}
                        &                                                         & Milo                                                                 & 35.80                    &                               \\ \cline{3-4}
                        &                                                         & EaTVul                                                               & \textbf{99.50}           &                               \\ \cline{2-5} 
                        & \multirow{3}{*}{LineVul}                                & \begin{tabular}[c]{@{}l@{}}Differentiable \\ Obfuscator\end{tabular} & 63.50                    & \multirow{3}{*}{83.45}        \\ \cline{3-4}
                        &                                                         & Milo                                                                 & 44.30                    &                               \\ \cline{3-4}
                        &                                                         & EaTVul                                                               & \textbf{92.30}           &                               \\ \hline
\multirow{7}{*}{CWE399} & \multirow{3}{*}{Asteria}                                & \begin{tabular}[c]{@{}l@{}}Differentiable \\ Obfuscator\end{tabular} & 58.80                    & \multirow{3}{*}{82.60}        \\ \cline{3-4}
                        &                                                         & Milo                                                                 & 26.30                    &                               \\ \cline{3-4}
                        &                                                         & EaTVul                                                               & \textbf{89.50}           &                               \\ \cline{2-5} 
                        & \multirow{3}{*}{LineVul}                                & \begin{tabular}[c]{@{}l@{}}Differentiable \\ Obfuscator\end{tabular} & 62.80                    & \multirow{3}{*}{83.50}        \\ \cline{3-4}
                        &                                                         & Milo                                                                 & 28.70                    &                               \\ \cline{3-4}
                        &                                                         & EaTVul                                                               & \textbf{89.20}           &                               \\ \hline
\multirow{7}{*}{CWE416} & \multirow{3}{*}{Asteria}                                & \begin{tabular}[c]{@{}l@{}}Differentiable \\ Obfuscator\end{tabular} & 52.50                    & \multirow{3}{*}{80.40}        \\ \cline{3-4}
                        &                                                         & Milo                                                                 & 20.35                    &                               \\ \cline{3-4}
                        &                                                         & EaTVul                                                               & \textbf{84.30}           &                               \\ \cline{2-5} 
                        & \multirow{3}{*}{LineVul}                                & \begin{tabular}[c]{@{}l@{}}Differentiable \\ Obfuscator\end{tabular} & 58.60                    & \multirow{3}{*}{81.70}        \\ \cline{3-4}
                        &                                                         & Milo                                                                 & 19.80                    &                               \\ \cline{3-4}
                        &                                                         & EaTVul                                                               & \textbf{87.50}           &                               \\ \hline
\end{tabular}
\label{tableQ5}
\end{table}

The results presented in the table showcase the effectiveness of the EaTVul attack strategy. Across various target models and datasets, the attack success rates achieved are generally high, indicating the strategy's ability to compromise the security of the models under attack.  These results suggest that the attack strategy is capable of effectively bypassing the defenses of the target models and exploiting vulnerabilities in the datasets. The high attack success rates across different scenarios and snippet sizes further emphasize the robustness and versatility of the strategy. These findings have significant implications for improving security measures and understanding potential vulnerabilities. By identifying weaknesses in the target models, researchers and practitioners can enhance their defense mechanisms and develop more resilient systems against similar attack strategies. Overall, the experimental results observed in the table demonstrate the efficacy and potential of the attack strategy in compromising the target models, shedding light on the need for further research and mitigation efforts in the field of cybersecurity.

\textit{\textbf{RQ5: How EaTVul behaves/performs regarding obfuscation/diversification methods?}}
To answer this question, we have conducted comparative experiments on two state-of-the-art target models with different vulnerability types. The target models include Asteria and LineVul because these two methods are representative techniques that leverage different ways of program representations and they were demonstrated to have high precision in detecting software vulnerabilities. For the data, we use three vulnerability types with different triggering mechanisms in the augmented Juliet C/C++ Test Suite Datasets. The baseline methods are Differentiable Obfuscator \cite{Srikant2021GeneratingAC} and Milo\cite{song2023milo}. To guarantee the optimal performance of baseline models, we set the perturbation strength to be 8.  And the perturbation strength represents the number of sites to transform.

Table \ref{tableQ5} displays the comparison results of evasion attacks on different datasets, CWE119, CWE399, and CWE416, using various target models (Asteria and LineVul) and attack models (our proposed EaTVul, and baseline obfuscation techniques such as Differentiable Obfuscator, Milo) with metrics such as Attack Success Rate (ASR) and F1-Score.
We can see that all the target models of Asteria \cite{wang2020combining} and LineVul demonstrate a high F1-Score of more than 80\%. Since the F1-Score reflects the balance between precision and recall, this means all the target models have fairly good performance before the evasion attack.
Results show that when targeting the Asteria model in the CWE119 dataset, Differentiable Obfuscator achieves an ASR of 66.40\%, indicating its success in evading the detection mechanisms of the Asteria model. In contrast, Milo has a lower ASR of 35.80\%, suggesting a lower success rate in evading detection. EaTVul, on the other hand, demonstrates a remarkably high ASR of 99.50\%, which corresponds to improvements of 33.10\% and 63.70\% over Differentiable Obfuscator and Milo, respectively, showcasing its effectiveness in successfully evading the Asteria model's detection mechanisms.
The same trend is observed when targeting the LineVul model within the CWE119 dataset, with Differentiable Obfuscator achieving a 63.50\% ASR, Milo achieving 44.30\%, and EaTVul achieving a high 92.30\% ASR. These results provide insights into the relative effectiveness of different attack models in evading detection on the specified dataset and target models. EaTVul stands out as particularly potent in achieving high ASR, suggesting its efficacy in crafting adversarial examples that successfully deceive the target models.
In the case of CWE399 and CWE416 datasets, EaTVul consistently shows high ASR values across different attack models, underscoring its effectiveness in evading detection. 

In summary, the outcomes demonstrate that our proposed EaTVul significantly enhances attack performance on state-of-the-art vulnerability detectors across various vulnerability types.

\begin{table}[!t]
\centering
\caption{Experiment results of ASR and F1-Score against two SOTA vulnerability detectors (i.e., FUNDED and VDet) on Java. ASR:Attack Success Rate (\%); F1-Score: F1-Score of Target Model Without Adversarial Samples (\%).}
\begin{tabular}{|l|l|l|c|}
\hline
Target Model            & Attack Model                                                         & ASR            & \multicolumn{1}{l|}{F1-Score} \\ \hline
\multirow{3}{*}{FUNDED} & \begin{tabular}[c]{@{}l@{}}Differentiable \\ Obfuscator\end{tabular} & 53.50          & \multirow{3}{*}{85.35}        \\ \cline{2-3}
                        & Milo                                                                 & 42.70          &                               \\ \cline{2-3}
                        & EaTVul                                                               & \textbf{88.60} &                               \\ \hline
\multirow{3}{*}{VDet}   & \begin{tabular}[c]{@{}l@{}}Differentiable \\ Obfuscator\end{tabular} & 63.50          & \multirow{3}{*}{86.60}        \\ \cline{2-3}
                        & Milo                                                                 & 62.80          &                               \\ \cline{2-3}
                        & EaTVul                                                               & \textbf{87.30} &                               \\ \hline
\end{tabular}
\label{tableQ6}
\end{table}

\textit{\textbf{RQ6: How effective is EaTVul when generalized to other programming languages?}}
To prove the generalizability of our proposed method, we extend the evaluation to two SOTA vulnerability detection models designed for Java programs. Specifically, we choose FUNDED \cite{wang2020combining} and VDet as the victim models \cite{mamede2022transformer}. Following the previous experimental settings, we compare two obfuscation techniques with our proposed method on the Java dataset \cite{wang2020combining}.

Table \ref{tableQ6} presents a comprehensive comparison of evasion attacks on two target models, FUNDED and VDet, using Differentiable Obfuscator, Milo, and our proposed EaTVul as attack models, with metrics including Attack Success Rate (ASR) and F1-Score.
We can see that all the target models of FUNDED \cite{wang2020combining} and VDet demonstrate a high F1-Score of 85.35\% and 86.60\%, this means all the target models have fairly good performance before the evasion attack.
Notably, our proposed method, EaTVul, consistently demonstrates superior performance in evasion attacks. 
Specifically, when FUNDED is chosen as the victim model, \textit{EaTVul} outperforms the Differentiable Obfuscator by 35.1\% and Milo by 45.9\% in terms of attack success rate. Likewise, EaTVul achieved much better attack performance than the Differentiable Obfuscator and Milo by 23.8\% and 24.5\% on VDet. 
This finding further substantiates the susceptibility of deep-learning-based vulnerability detectors to adversarial attacks, especially when considering semantic information alone, emphasizing the need to incorporate additional control flow and data flow information for method resilience.

In summary, the experimental outcomes with the Java dataset substantiate the broad applicability of our proposed method in executing evasion attacks within real-world project contexts.

\subsection{Limitations}
EaTVul system has several limitations. We plan to address the following issues in the future.

This study focuses on evasion attacks. From a defence perspective, it would be great to automatically recognize an adversarial sample and tell if it is actually an adversarial sample or not. However, this is a research topic in itself, and this topic is outside the scope of this work at the current stage. We believe the effectiveness of evasion attacks can be influenced by defense mechanisms. If the vulnerability detection system implements robust defense techniques such as input sanitization, anomaly detection, or ensemble models, the success rate of evasion attacks may significantly decrease. We leave this as future work.

The proposed EaTVul relies on the important samples (i.e., support vectors) identified by SVM. In this case, the number of important samples is limited to the number of support vectors. We would like to explore other methods, such as information retrieval \cite{liu2019data} method, to further identify important samples based on the probability of being non-vulnerable and to enlarge the sample space in the next work.

EaTVul is currently limited to utilizing an attention model to identify the most important features. However, our future plans involve incorporating other interpretation models such as LIME and SHAP \cite{slack2020fooling, sharma2022explainable} and exploring the use of these approaches. By incorporating these techniques, we aim to enhance the effectiveness and versatility of EaTVul in evading machine learning-based vulnerability detection systems.

In this study, we employ a fuzzy genetic algorithm to select attack samples and enhance the success rate. However, the applicability of FGA is constrained by the inherent attacking effectiveness of the constructed adversarial code snippets. FGA primarily endeavors to hunt for the optimal combination of pre-existing fragments, a process significantly influenced by the quality of the adversarial code snippets, particularly their naturalness. In the future, we plan to explore other large language models to automatically generate adversarial samples to launch an attack

Moreover, this study assumes attackers have knowledge of training data, which may not always be true in real-world scenarios. However, there are studies showing that one can get knowledge of the training data by using recently developed technologies \cite{lee2022query, hu2023generating}, therefore, it would be interesting to explore the training data by considering these technologies. 

EaTVul has been focused on the scenario of adversarial learning at source code level software vulnerability detection. However, software vulnerability detection at the binary level is a hot topic recently, and we plan to further apply and develop our proposed scheme to binary code in the future. Especially, the comparison with adversarial malware detection in binaries will be an interesting research direction.

In this study, EaTVul has restricted the size of the adversarial data to less than 8 lines to ensure its concealment based on our historical study. In future research, we aim to investigate this further by incorporating knowledge from natural language processing \cite{qiu2022adversarial}. Our expectation is that the size of the adversarial data can be further optimized and reduced.

\section{Conclusion}
In conclusion, the rapid advancement of technology and the increasing complexity of cyber threats have made cybersecurity a critical concern in today's digital world. 
While deep learning techniques have shown promise in detecting vulnerabilities and improving the accuracy of software vulnerability detection systems, they are not immune to attacks themselves. Adversarial examples can exploit vulnerabilities in deep neural networks, compromising the security of the entire system.

This paper has addressed the issue of adversarial attacks on machine learning-based software vulnerability detection systems. It has introduced a novel attack strategy called EaTVul, which successfully generates adversarial examples to evade detection. The proposed strategy utilizes support vector machines, attention mechanisms, chatGPT, and a fuzzy genetic algorithm to identify important samples, key words, generate noise data, and select optimal seed adversarial data. Extensive experiments have demonstrated the efficacy of EaTVul in bypassing machine learning detection systems and altering the prediction outcomes.



\bibliographystyle{plain}
\bibliography{reference_svd}

\begin{thebibliography}{10}

\bibitem{chan2023transformer}
Aaron Chan, Anant Kharkar, Roshanak~Zilouchian Moghaddam, Yevhen Mohylevskyy, Alec Helyar, Eslam Kamal, Mohamed Elkamhawy, and Neel Sundaresan.
\newblock Transformer-based vulnerability detection in code at edittime: Zero-shot, few-shot, or fine-tuning?
\newblock {\em arXiv preprint arXiv:2306.01754}, 2023.

\bibitem{christou2023ivysyn}
Neophytos Christou, Di~Jin, Vaggelis Atlidakis, Baishakhi Ray, and Vasileios~P Kemerlis.
\newblock $\{$IvySyn$\}$: Automated vulnerability discovery in deep learning frameworks.
\newblock In {\em 32nd USENIX Security Symposium (USENIX Security 23)}, pages 2383--2400, 2023.

\bibitem{feng2020efficient}
Hantao Feng, Xiaotong Fu, Hongyu Sun, He~Wang, and Yuqing Zhang.
\newblock Efficient vulnerability detection based on abstract syntax tree and deep learning.
\newblock In {\em IEEE INFOCOM 2020-IEEE Conference on Computer Communications Workshops (INFOCOM WKSHPS)}, pages 722--727. IEEE, 2020.

\bibitem{fu2022linevul}
Michael Fu and Chakkrit Tantithamthavorn.
\newblock Linevul: a transformer-based line-level vulnerability prediction.
\newblock In {\em Proceedings of the 19th International Conference on Mining Software Repositories}, pages 608--620, 2022.

\bibitem{garg-ramakrishnan-2020-bae}
Siddhant Garg and Goutham Ramakrishnan.
\newblock {BAE}: {BERT}-based adversarial examples for text classification.
\newblock In Bonnie Webber, Trevor Cohn, Yulan He, and Yang Liu, editors, {\em Proceedings of the 2020 Conference on Empirical Methods in Natural Language Processing (EMNLP)}, pages 6174--6181, Online, November 2020. Association for Computational Linguistics.

\bibitem{ghaffarian2017software}
Seyed~Mohammad Ghaffarian and Hamid~Reza Shahriari.
\newblock Software vulnerability analysis and discovery using machine-learning and data-mining techniques: A survey.
\newblock {\em ACM Computing Surveys (CSUR)}, 50(4):1--36, 2017.

\bibitem{ramakrishnan2022semantic}
Jordan Henke, Goutham Ramakrishnan, Zi~Wang, Aws Albarghouth, Somesh Jha, and Thomas Reps.
\newblock Semantic robustness of models of source code.
\newblock In {\em 2022 IEEE International Conference on Software Analysis, Evolution and Reengineering (SANER)}, pages 526--537. IEEE, 2022.

\bibitem{hin2022linevd}
David Hin, Andrey Kan, Huaming Chen, and M~Ali Babar.
\newblock Linevd: Statement-level vulnerability detection using graph neural networks.
\newblock In {\em Proceedings of the 19th International Conference on Mining Software Repositories}, pages 596--607, 2022.

\bibitem{hu2023generating}
Weiwei Hu and Ying Tan.
\newblock Generating adversarial malware examples for black-box attacks based on gan.
\newblock In {\em Data Mining and Big Data: 7th International Conference, DMBD 2022, Beijing, China, November 21--24, 2022, Proceedings, Part II}, pages 409--423. Springer, 2023.

\bibitem{jain2023code}
Ridhi Jain, Nicole Gervasoni, Mthandazo Ndhlovu, and Sanjay Rawat.
\newblock A code centric evaluation of c/c++ vulnerability datasets for deep learning based vulnerability detection techniques.
\newblock In {\em Proceedings of the 16th Innovations in Software Engineering Conference}, pages 1--10, 2023.

\bibitem{jin2020bert}
Di~Jin, Zhijing Jin, Joey~Tianyi Zhou, and Peter Szolovits.
\newblock Is bert really robust? a strong baseline for natural language attack on text classification and entailment.
\newblock In {\em Proceedings of the AAAI conference on artificial intelligence}, volume~34, pages 8018--8025, 2020.

\bibitem{kolosnjaji2018adversarial}
Bojan Kolosnjaji, Ambra Demontis, Battista Biggio, Davide Maiorca, Giorgio Giacinto, Claudia Eckert, and Fabio Roli.
\newblock Adversarial malware binaries: Evading deep learning for malware detection in executables.
\newblock In {\em 2018 26th European signal processing conference (EUSIPCO)}, pages 533--537. IEEE, 2018.

\bibitem{le2022survey}
Triet~HM Le, Huaming Chen, and M~Ali Babar.
\newblock A survey on data-driven software vulnerability assessment and prioritization.
\newblock {\em ACM Computing Surveys}, 55(5):1--39, 2022.

\bibitem{lee2022query}
Deokjae Lee, Seungyong Moon, Junhyeok Lee, and Hyun~Oh Song.
\newblock Query-efficient and scalable black-box adversarial attacks on discrete sequential data via bayesian optimization.
\newblock In {\em International Conference on Machine Learning}, pages 12478--12497. PMLR, 2022.

\bibitem{li2021arms}
Deqiang Li, Qianmu Li, Yanfang Ye, and Shouhuai Xu.
\newblock Arms race in adversarial malware detection: A survey.
\newblock {\em ACM Computing Surveys (CSUR)}, 55(1):1--35, 2021.

\bibitem{li2021conaml}
Jiangnan Li, Yingyuan Yang, Jinyuan~Stella Sun, Kevin Tomsovic, and Hairong Qi.
\newblock Conaml: Constrained adversarial machine learning for cyber-physical systems.
\newblock In {\em Proceedings of the 2021 ACM Asia Conference on Computer and Communications Security}, pages 52--66, 2021.

\bibitem{li-etal-2020-bert-attack}
Linyang Li, Ruotian Ma, Qipeng Guo, Xiangyang Xue, and Xipeng Qiu.
\newblock {BERT}-{ATTACK}: Adversarial attack against {BERT} using {BERT}.
\newblock In Bonnie Webber, Trevor Cohn, Yulan He, and Yang Liu, editors, {\em Proceedings of the 2020 Conference on Empirical Methods in Natural Language Processing (EMNLP)}, pages 6193--6202, Online, November 2020. Association for Computational Linguistics.

\bibitem{li2021pdgraph}
Qiang Li, Jinke Song, Dawei Tan, Haining Wang, and Jiqiang Liu.
\newblock Pdgraph: a large-scale empirical study on project dependency of security vulnerabilities.
\newblock In {\em 2021 51st Annual IEEE/IFIP International Conference on Dependable Systems and Networks (DSN)}, pages 161--173. IEEE, 2021.

\bibitem{li2016vulpecker}
Zhen Li, Deqing Zou, Shouhuai Xu, Hai Jin, Hanchao Qi, and Jie Hu.
\newblock Vulpecker: an automated vulnerability detection system based on code similarity analysis.
\newblock In {\em Proceedings of the 32nd annual conference on computer security applications}, pages 201--213, 2016.

\bibitem{li2018vuldeepecker}
Zhen Li, Deqing Zou, Shouhuai Xu, Xinyu Ou, Hai Jin, Sujuan Wang, Zhijun Deng, and Yuyi Zhong.
\newblock Vuldeepecker: A deep learning-based system for vulnerability detection.
\newblock {\em in 25th Annual Network and Distributed System Security Symposium (NDSS 2018), San Diego, California, USA, February 18- 21, 2018 (EI/CCF-A)}, 2018.

\bibitem{lin2020software}
Guanjun Lin, Sheng Wen, Qing-Long Han, Jun Zhang, and Yang Xiang.
\newblock Software vulnerability detection using deep neural networks: a survey.
\newblock {\em Proceedings of the IEEE}, 108(10):1825--1848, 2020.

\bibitem{lin2020deep}
Guanjun Lin, Wei Xiao, Jun Zhang, and Yang Xiang.
\newblock Deep learning-based vulnerable function detection: A benchmark.
\newblock In {\em Information and Communications Security: 21st International Conference, ICICS 2019, Beijing, China, December 15--17, 2019, Revised Selected Papers 21}, pages 219--232. Springer, 2020.

\bibitem{lin2017poster}
Guanjun Lin, Jun Zhang, Wei Luo, Lei Pan, and Yang Xiang.
\newblock Poster: Vulnerability discovery with function representation learning from unlabeled projects.
\newblock In {\em Proceedings of the 2017 ACM SIGSAC Conference on Computer and Communications Security}, pages 2539--2541, 2017.

\bibitem{liu2019data}
Jiaying Liu, Xiangjie Kong, Xinyu Zhou, Lei Wang, Da~Zhang, Ivan Lee, Bo~Xu, and Feng Xia.
\newblock Data mining and information retrieval in the 21st century: A bibliographic review.
\newblock {\em Computer science review}, 34:100193, 2019.

\bibitem{liu2019cyber}
Shigang Liu, Mahdi Dibaei, Yonghang Tai, Chao Chen, Jun Zhang, and Yang Xiang.
\newblock Cyber vulnerability intelligence for internet of things binary.
\newblock {\em IEEE Transactions on Industrial Informatics}, 16(3):2154--2163, 2019.

\bibitem{liu2018data}
Shigang Liu, Jun Zhang, Yu~Wang, Wanlei Zhou, Yang Xiang, and Olivier~De Vel.
\newblock A data-driven attack against support vectors of svm.
\newblock In {\em Proceedings of the 2018 on Asia Conference on Computer and Communications Security}, pages 723--734, 2018.

\bibitem{Luo0WTXZLL23}
Zhenhao Luo, Pengfei Wang, Baosheng Wang, Yong Tang, Wei Xie, Xu~Zhou, Danjun Liu, and Kai Lu.
\newblock Vulhawk: Cross-architecture vulnerability detection with entropy-based binary code search.
\newblock In {\em 30th Annual Network and Distributed System Security Symposium, {NDSS} 2023, San Diego, California, USA, February 27 - March 3, 2023}. The Internet Society, 2023.

\bibitem{machado2021adversarial}
Gabriel~Resende Machado, Eug{\^e}nio Silva, and Ronaldo~Ribeiro Goldschmidt.
\newblock Adversarial machine learning in image classification: A survey toward the defender’s perspective.
\newblock {\em ACM Computing Surveys (CSUR)}, 55(1):1--38, 2021.

\bibitem{maiorca2019towards}
Davide Maiorca, Battista Biggio, and Giorgio Giacinto.
\newblock Towards adversarial malware detection: Lessons learned from pdf-based attacks.
\newblock {\em ACM Computing Surveys (CSUR)}, 52(4):1--36, 2019.

\bibitem{mamede2022transformer}
Cl{\'a}udia Mamede, Eduard Pinconschi, and Rui Abreu.
\newblock A transformer-based ide plugin for vulnerability detection.
\newblock In {\em Proceedings of the 37th IEEE/ACM International Conference on Automated Software Engineering}, pages 1--4, 2022.

\bibitem{mirsky2023vulchecker}
Yisroel Mirsky, George Macon, Michael Brown, Carter Yagemann, Matthew Pruett, Evan Downing, Sukarno Mertoguno, and Wenke Lee.
\newblock Vulchecker: Graph-based vulnerability localization in source code.
\newblock In {\em 31st USENIX Security Symposium, Security 2022}, 2023.

\bibitem{miyamoto2008algorithms}
Sadaaki Miyamoto, Hodetomo Ichihashi, Katsuhiro Honda, and Hidetomo Ichihashi.
\newblock {\em Algorithms for fuzzy clustering}, volume~10.
\newblock Springer, 2008.

\bibitem{pearce2023examining}
Hammond Pearce, Benjamin Tan, Baleegh Ahmad, Ramesh Karri, and Brendan Dolan-Gavitt.
\newblock Examining zero-shot vulnerability repair with large language models.
\newblock In {\em 2023 IEEE Symposium on Security and Privacy (SP)}, pages 2339--2356. IEEE, 2023.

\bibitem{pisner2020support}
Derek~A Pisner and David~M Schnyer.
\newblock Support vector machine.
\newblock In {\em Machine learning}, pages 101--121. Elsevier, 2020.

\bibitem{qiu2022adversarial}
Shilin Qiu, Qihe Liu, Shijie Zhou, and Wen Huang.
\newblock Adversarial attack and defense technologies in natural language processing: A survey.
\newblock {\em Neurocomputing}, 492:278--307, 2022.

\bibitem{ramakrishnan2022backdoors}
Goutham Ramakrishnan and Aws Albarghouthi.
\newblock Backdoors in neural models of source code.
\newblock In {\em 2022 26th International Conference on Pattern Recognition (ICPR)}, pages 2892--2899. IEEE, 2022.

\bibitem{rosenberg2021adversarial}
Ishai Rosenberg, Asaf Shabtai, Yuval Elovici, and Lior Rokach.
\newblock Adversarial machine learning attacks and defense methods in the cyber security domain.
\newblock {\em ACM Computing Surveys (CSUR)}, 54(5):1--36, 2021.

\bibitem{schuster2021you}
Roei Schuster, Congzheng Song, Eran Tromer, and Vitaly Shmatikov.
\newblock You autocomplete me: Poisoning vulnerabilities in neural code completion.
\newblock In {\em 30th USENIX Security Symposium (USENIX Security 21)}, pages 1559--1575, 2021.

\bibitem{sendner2023smarter}
Christoph Sendner, Huili Chen, Hossein Fereidooni, Lukas Petzi, Jan K{\"o}nig, Jasper Stang, Alexandra Dmitrienko, Ahmad-Reza Sadeghi, and Farinaz Koushanfar.
\newblock Smarter contracts: Detecting vulnerabilities in smart contracts with deep transfer learning.
\newblock In {\em To appear at the Network and Distributed System Security Symposium (NDSS)}, 2023.

\bibitem{sharma2022explainable}
Deepak~Kumar Sharma, Jahanavi Mishra, Aeshit Singh, Raghav Govil, Gautam Srivastava, and Jerry Chun-Wei Lin.
\newblock Explainable artificial intelligence for cybersecurity.
\newblock {\em Computers and Electrical Engineering}, 103:108356, 2022.

\bibitem{slack2020fooling}
Dylan Slack, Sophie Hilgard, Emily Jia, Sameer Singh, and Himabindu Lakkaraju.
\newblock Fooling lime and shap: Adversarial attacks on post hoc explanation methods.
\newblock In {\em Proceedings of the AAAI/ACM Conference on AI, Ethics, and Society}, pages 180--186, 2020.

\bibitem{song2023milo}
Leo Song and Steven~HH Ding.
\newblock Milo: Attacking deep pre-trained model for programming languages tasks with anti-analysis code obfuscation.
\newblock In {\em 2023 IEEE 47th Annual Computers, Software, and Applications Conference (COMPSAC)}, pages 586--594. IEEE, 2023.

\bibitem{Srikant2021GeneratingAC}
Shashank Srikant, Sijia Liu, Tamara Mitrovska, Shiyu Chang, Quanfu Fan, Gaoyuan Zhang, and Una-May O’Reilly.
\newblock Generating adversarial computer programs using optimized obfuscations.
\newblock {\em In International Conference on Learning Representations (ICLR)}, 2021.

\bibitem{van2008visualizing}
Laurens Van~der Maaten and Geoffrey Hinton.
\newblock Visualizing data using t-sne.
\newblock {\em Journal of machine learning research}, 9(11), 2008.

\bibitem{vaswani2017attention}
Ashish Vaswani, Noam Shazeer, Niki Parmar, Jakob Uszkoreit, Llion Jones, Aidan~N Gomez, {\L}ukasz Kaiser, and Illia Polosukhin.
\newblock Attention is all you need.
\newblock {\em Advances in neural information processing systems}, 30, 2017.

\bibitem{wang2020combining}
Huanting Wang, Guixin Ye, Zhanyong Tang, Shin~Hwei Tan, Songfang Huang, Dingyi Fang, Yansong Feng, Lizhong Bian, and Zheng Wang.
\newblock Combining graph-based learning with automated data collection for code vulnerability detection.
\newblock {\em IEEE Transactions on Information Forensics and Security}, 16:1943--1958, 2020.

\bibitem{wu2020multitasking}
Dongrui Wu and Xianfeng Tan.
\newblock Multitasking genetic algorithm (mtga) for fuzzy system optimization.
\newblock {\em IEEE Transactions on Fuzzy Systems}, 28(6):1050--1061, 2020.

\bibitem{yan2022survey}
Senming Yan, Jing Ren, Wei Wang, Limin Sun, Wei Zhang, and Quan Yu.
\newblock A survey of adversarial attack and defense methods for malware classification in cyber security.
\newblock {\em ACM Computing Surveys(CSUR)}, 25(1):467--496, 2023.

\bibitem{yang2020greedy}
Puyudi Yang, Jianbo Chen, Cho-Jui Hsieh, Jane-Ling Wang, and Michael~I Jordan.
\newblock Greedy attack and gumbel attack: Generating adversarial examples for discrete data.
\newblock {\em The Journal of Machine Learning Research}, 21(1):1613--1648, 2020.

\bibitem{yang2021asteria}
Shouguo Yang, Long Cheng, Yicheng Zeng, Zhe Lang, Hongsong Zhu, and Zhiqiang Shi.
\newblock Asteria: Deep learning-based ast-encoding for cross-platform binary code similarity detection.
\newblock In {\em 2021 51st Annual IEEE/IFIP International Conference on Dependable Systems and Networks (DSN)}, pages 224--236. IEEE, 2021.

\bibitem{yang2022natural}
Zhou Yang, Jieke Shi, Junda He, and David Lo.
\newblock Natural attack for pre-trained models of code.
\newblock In {\em Proceedings of the 44th International Conference on Software Engineering}, pages 1482--1493, 2022.

\bibitem{yefet2020adversarial}
Noam Yefet, Uri Alon, and Eran Yahav.
\newblock Adversarial examples for models of code.
\newblock {\em Proceedings of the ACM on Programming Languages}, 4(OOPSLA):1--30, 2020.

\bibitem{yu2023advulcode}
Xueqi Yu, Zhen Li, Xiang Huang, and Shasha Zhao.
\newblock Advulcode: Generating adversarial vulnerable code against deep learning-based vulnerability detectors.
\newblock {\em Electronics}, 12(4):936, 2023.

\bibitem{zeng2021}
Guoyang Zeng, Fanchao Qi, Qianrui Zhou, Tingji Zhang, Zixian Ma, Bairu Hou, Yuan Zang, Zhiyuan Liu, and Maosong Sun.
\newblock Openattack: An open-source textual adversarial attack toolkit.
\newblock In {\em In Proceedings of the 59th Annual Meeting of the Association for Computational Linguistics and the 11th International Joint Conference on Natural Language Processing: System Demonstrations}, pages 363--371, 01 2021.

\bibitem{zhang2020generating}
Huangzhao Zhang, Zhuo Li, Ge~Li, Lei Ma, Yang Liu, and Zhi Jin.
\newblock Generating adversarial examples for holding robustness of source code processing models.
\newblock In {\em Proceedings of the AAAI Conference on Artificial Intelligence}, volume~34, pages 1169--1176, 2020.

\bibitem{zhou2022adversarial}
Yu~Zhou, Xiaoqing Zhang, Juanjuan Shen, Tingting Han, Taolue Chen, and Harald Gall.
\newblock Adversarial robustness of deep code comment generation.
\newblock {\em ACM Transactions on Software Engineering and Methodology (TOSEM)}, 31(4):1--30, 2022.

\end{thebibliography}

\end{document}